\documentclass[aps,prb,amsmath,amssymb,amsfont,showpacs,superscriptaddress,reprint]{revtex4-1}
\usepackage{amsmath}
\usepackage{xcolor}
\usepackage{braket}
\usepackage[artemisia]{textgreek}
\usepackage{graphicx}

\begin{document}

\title{Electronic and magnetic properties of black phosphorus}

\author{Bence G. M\'{a}rkus}
\affiliation{Department of Physics, Budapest University of Technology and Economics and MTA-BME Lend\"{u}let Spintronics Research Group (PROSPIN), P.O. Box 91, H-1521 Budapest, Hungary}

\author{Ferenc Simon}
\affiliation{Department of Physics, Budapest University of Technology and Economics and MTA-BME Lend\"{u}let Spintronics Research Group (PROSPIN), P.O. Box 91, H-1521 Budapest, Hungary}

\author{K\'{a}roly Nagy}
\affiliation{Department of Physics, Budapest University of Technology and Economics and MTA-BME Lend\"{u}let Spintronics Research Group (PROSPIN), P.O. Box 91, H-1521 Budapest, Hungary}

\author{Titusz Feh\'{e}r}
\affiliation{Department of Physics, Budapest University of Technology and Economics and MTA-BME Lend\"{u}let Spintronics Research Group (PROSPIN), P.O. Box 91, H-1521 Budapest, Hungary}

\author{Stefan Wild}
\affiliation{Department of Chemistry and Pharmacy Friedrich-Alexander-Universit\"{a}t Erlangen-N\"{u}rnberg (FAU) D-91054 Erlangen, Germany and Joint Institute of Advanced Materials and Processes (ZMP), D-90762 F\"{u}rth, Germany}

\author{Gonzalo Abell\'{a}n}
\affiliation{Department of Chemistry and Pharmacy Friedrich-Alexander-Universit\"{a}t Erlangen-N\"{u}rnberg (FAU) D-91054 Erlangen, Germany and Joint Institute of Advanced Materials and Processes (ZMP), D-90762 F\"{u}rth, Germany}

\author{Julio C. Chac\'{o}n-Torres}
\affiliation{Yachay Tech. University, School of Physical Sciences and Nanotechnology, 100119 Urcuqu\'{i}, Ecuador}
\affiliation{Institut f\"{u}r Experimental Physik, Freie Universit\"{a}t Berlin, Arnimallee 14, D-14195 Berlin, Germany}

\author{Andreas Hirsch}
\affiliation{Department of Chemistry and Pharmacy Friedrich-Alexander-Universit\"{a}t Erlangen-N\"{u}rnberg (FAU) D-91054 Erlangen, Germany and Joint Institute of Advanced Materials and Processes (ZMP), D-90762 F\"{u}rth, Germany}

\author{Frank Hauke}
\affiliation{Department of Chemistry and Pharmacy Friedrich-Alexander-Universit\"{a}t Erlangen-N\"{u}rnberg (FAU) D-91054 Erlangen, Germany and Joint Institute of Advanced Materials and Processes (ZMP), D-90762 F\"{u}rth, Germany}


\begin{abstract}Black phosphorus has emerged as the next member in the graphene inspired two-dimensional materials family. Its electronic and magnetic properties are studied herein using electron and nuclear magnetic resonance techniques (ESR and NMR) and microwave conductivity measurement. The latter is a unique technique to study conductivity on air sensitive samples. The ESR study indicates the absence of free charge carriers and no sign of paramagnetic defects are found.$^{31}$P NMR shows the presence of a characteristic Pake doublet structure due to the interaction between $I=1/2$ nuclei. Microwave conductivity shows, in accordance with the ESR results, that black phosphorus behaves as a semiconductor and we identify extrinsic and intrinsic charge carrier contributions to the conductivity and extracted the sizes of the gaps. ESR measurement also yields that bP might find applications as a microwave absorbent.\end{abstract}

\maketitle

\section{Introduction}

The discovery of graphene \cite{novoselov2004} opened up a new perspective for the application and fundamental studies of two-dimensional materials. Black phosphorus was rediscovered as a potentially interesting 2D material given that it has been known for a century \cite{bridgman1914,Liu2014,ling2015,gusmao2017}. While structurally similar, the most important electronic difference compared to graphene is the expected small gap semiconducting behavior which would make this material to lie between the large gap transition metal dichalcogenides and graphene \cite{castellanos-gomez2015}. This has clear implications on the utility of black phosphorus in electronic devices.

Numerous efforts are made to exfoliate single layer or few layer \cite{Li2014,Lu2014,Castellanos-Gomez2014,Koening2014,Buscema2014,zhang2014,brent2014,yasaei2015,kang2015,hanlon2015,Abellan2016} black phosphorus (bP) from bulk crystals, yet production of good quality single layer phosphorene in high yields remain an open question. An important hindrance is that while the bulk material is more or less stable in air (surface oxidation can passivate the material if water is not present \cite{edmonds2015}), the mono- and few-layer flakes are unstable in the presence of oxygen or moisture (due to exoenergetic chemisorption of oxygen \cite{ziletti2015}) resulting phosphorus oxides and acids \cite{gusmao2017,hanlon2015,edmonds2015,ziletti2015,favron2015,wood2014,ziletti2015_2,wang2016_1,wang2016_2,huang2016}. Much as the air and water sensitivity makes the studies difficult, it is expected that bP would serve as a starting material for a new class of hybrids with plenty of applications, much like graphite is the mother compound of graphite intercalation compounds \cite{Dresselhaus1981}, which turned out to be rewarding for both fundamental science and applications.

In order to proceed with the hybrid preparation it is imperative to characterize bP as well as possible. Concerning the electronic properties, magnetic resonance and conductivity measurements can provide additional information. Transport and photoelectron spectroscopy studies \cite{Keyes1953,Warschauer1963,Maruyama1981,Harada1982,Akahama1983,Narita1983,Takahashi1984,Hayashi1984,Baba1991_1,Han2014} indicate the presence of a moderate band gap of about $0.3$ eV for bulk, which can be increased up to $2$ eV with reducing the number of layers \cite{Liu2014,Li2014,Koening2014}. Electron spin resonance (ESR) spectroscopy can indicate if localized or delocalized unpaired electrons are present and also hint at the presence of impurities, nuclear magnetic resonance (NMR) is a powerful tool to characterize the chemical environment of the studied nuclei and to tell whether any fluctuating electron spins are present in the vicinity of the nucleus (either localized or itinerant). The $^{31}$P NMR spectra was also studied using magic angle spinning (MAS) \cite{wang2016_1,Lange2007}, paying a special attention on the degradation of the material. In addition, microwave conductivity allows a direct and contactless measurement of the conduction properties which is particularly suited for air sensitive materials where the conventional contact methods are impractical. Clearly a study with these techniques is desired due to the wealth of informations they could provide.

With this motivation in mind, we studied bP using magnetic resonance (ESR and NMR) and conductivity measurements. ESR indicates the absence of free charge carriers and shows a negligible amount of impurities only. $^{31}$P NMR indicates a semiconducting/insulating behavior with the well-known Pake doublet structure which is characteristic for $I=1/2$ nuclei, which interact through the magnetic dipole-dipole interaction. The temperature dependent microwave conductivity showed that the material is semiconducting with two features (large and small gap). This double activated behavior is well known for weakly doped semiconductors.

\section{Experimental}

Black phosphorus was obtained from smart-elements with purity of $99.998\%$. The commercial bP crystals were grounded inside an Argon-filled glovebox (MBraun, $<0.1$ ppm of O$_2$ and H$_2$O) and used as a crushed powder. For the NMR and ESR experiments $52$ mg, for microwave conductivity $2$ mg of bP was put into quartz tubes, evacuated to high vacuum ($2\times10^{-6}$ mbar) and sealed under $20$ mbar of He (purity $99.9999\%$). The aqueous H$_3$PO$_4$ was nominally $85$ wt$\%$ and obtained from Hungarian Fine Chemicals ($98\%$ pure), and was kept in a sealed teflon holder to avoid moisture uptake from air. Phosphorus pentoxide, P$_2$O$_5$ was from Merck with purity of $98\%$ . Approximately $30$ mg of powder was put into quartz tube and dried under $2\times10^{-6}$ mbar of pressure at $120~^{\circ}$C for $2$ hours to evaporate the transformed material, then sealed.

Raman spectra were acquired on a commercial LabRam HR Evolution confocal Raman microscope (Horiba) equipped with an automated XYZ table and a laser spot size of  $1$ {\textmu}m (Olympus LMPlanFl $100$, $\mathrm{NA} = 0.80$ ). All measurements were conducted in backscattering geometry using an excitation wavelength of $532$ nm, with an acquisition time of $2$ s and a grating of $1800$ grooves$/$mm. To minimize the photo-induced laser oxidation of the samples, the laser intensity was kept at $0.88$ mW. The scattered photons were detected with a cooled CCD, the sample was placed on a Si/SiO$_2$ wafer.

Static NMR measurements were carried out on a commercial Bruker Avance 300 instrument with a custom built, tunable solenoid probehead. The DC magnetic field is $7$ T, the corresponding $^{31}$P frequency is about $121$ MHz. The NMR shift was measured with respect to the H$_3$PO$_4$ reference sample. The $T_2$ transversal or spin-spin relaxation time was determined from separate spin echo signals with varying delay between a $90$ and $180$ degree pulses (varied from $20$ {\textmu}s to $100$ {\textmu}s). $T_1$ longitudinal or spin-lattice relaxation time was determined from a free induced decay (FID) signal with different repetition times (varied from $100$ ms to $15$ s) \cite{slichter,fukushima}. For ESR measurements a Bruker Elexsys E500 $X$-band spectrometer was used. Both the NMR and ESR measurements were done at room temperature.

Microwave conductivity measurements were done with the cavity perturbation technique \cite{Klein1993,Donovan1993}. The method relies on placing the sample in a microwave cavity (into the node of the microwave electric, and maximum of the microwave magnetic field), where the alternating microwave magnetic field induces eddy current in the sample. This configuration is very sensitive to small changes in the sample conductivity for fine grains samples \cite{Kitano2002}. A measurement of resistivity, $\rho$, proceeds by detecting the change in the loaded cavity quality factor ($Q_{\text{L}}$), which is corrected by measuring the $Q$ of the unloaded cavity ($Q_0$). It is known that for fine grains $\rho \propto \left(1/Q_{\text{L}}-1/Q_0\right)^{-1}$. A limitation of the method is that it cannot provide absolute values of the resistivity as sample size distribution related correction would be needed. An advantage of the method is that it is very sensitive to small changes in the relative value of $\rho$ and that it can be readily applied for air sensitive sample and for fine powders for which the conventional, contact based, methods are impractical or impossible. The $Q$ factor of the cavity is measured via rapid frequency sweeps near the resonance. A fit to the obtained resonance curve yields the position, $\omega_0$, and width, $\rm{\Delta} \omega$, of the resonance. $Q$ is obtained from $\omega_0/\rm{\Delta} \omega$.

\section{Results}

In bP direct measurements are hindered by the extreme sensitivity of the material to air, moisture and visible light. Thus, one has to apply contactless methods such as NMR, ESR and microwave conductivity to get a deeper insight of the electronic and magnetic properties.

\subsection{Raman spectroscopy}

\begin{figure}[h!]
\includegraphics*[width=0.9\linewidth]{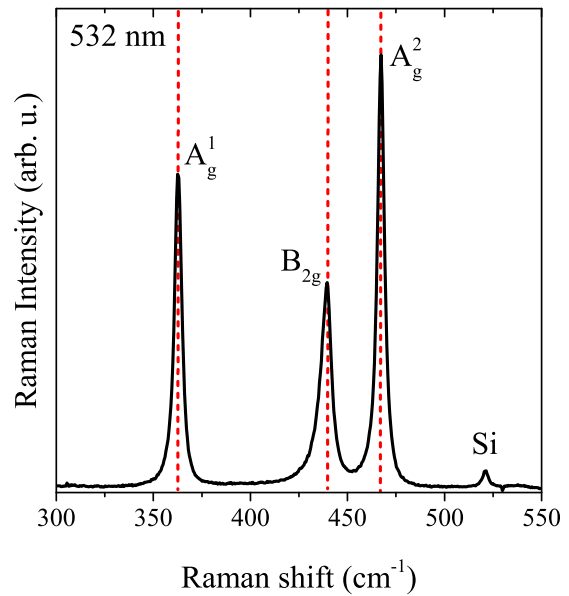}
\caption{Raman spectrum of black phosphorus, dashed lines indicate the position of the characteristic A$_{\text{g}}^1$, B$_{2\text{g}}$ and A$_{\text{g}}^2$ Raman peaks, whose positions agrees well with the literature values \cite{zhang2014,Sugai1985,Vanderborgh1989,Akahama1997,Late2015,Riberio2015,Ling2016}.}
\label{fig:raman}
\end{figure}

To verify the quality of the used bP material, first the Raman spectrum was recorded. Fig. \ref{fig:raman}. shows the spectrum obtained at $532$ nm excitation wavelength.

From the spectrum the $3$ characteristic features: A$_{\text{g}}^1$ at $363(1)$ cm$^{-1}$, B$_{2\text{g}}$ at $440(1)$ cm$^{-1}$ and A$_{\text{g}}^2$ at $467(1)$ cm$^{-1}$ modes of bP were identified. The position of the peaks agree well with literature results \cite{zhang2014,Sugai1985,Vanderborgh1989,Akahama1997,Late2015,Riberio2015,Ling2016}, thus the grade of the examined material is satisfying.

\subsection{NMR}

Early studies of black phosphorus were hindered by the rapid degradation of the material under ambient conditions due to its remarked oxophilicity \cite{huang2016}. Further studies showed that the degradation is due to oxidation and in the presence of water the oxides can transform into acids. In excess oxygen not just P$-$O and P$=$O formation is present in bP, but side products also appear as separate molecules. Using \emph{a priori} knowledge from white phosphorus one expects that only PO$_3$ is formed at room temperature, which can turn into phosphorus acid, H$_3$PO$_3$ \cite{greenwood1998}. However, recent experiments showed that higher oxidation level of phosphorus is also present, namely phosphorus pentoxide P$_2$O$_5$ and phosphoric acid H$_3$PO$_4$ \cite{hanlon2015,edmonds2015,ziletti2015,favron2015,wood2014,wang2016_1,wang2016_2,huang2016}. Most probably these reactions occur because the first step of the oxidation is so exothermic that it can activate the second reaction channel as well.

\begin{figure}[h!]
\includegraphics*[width=\linewidth]{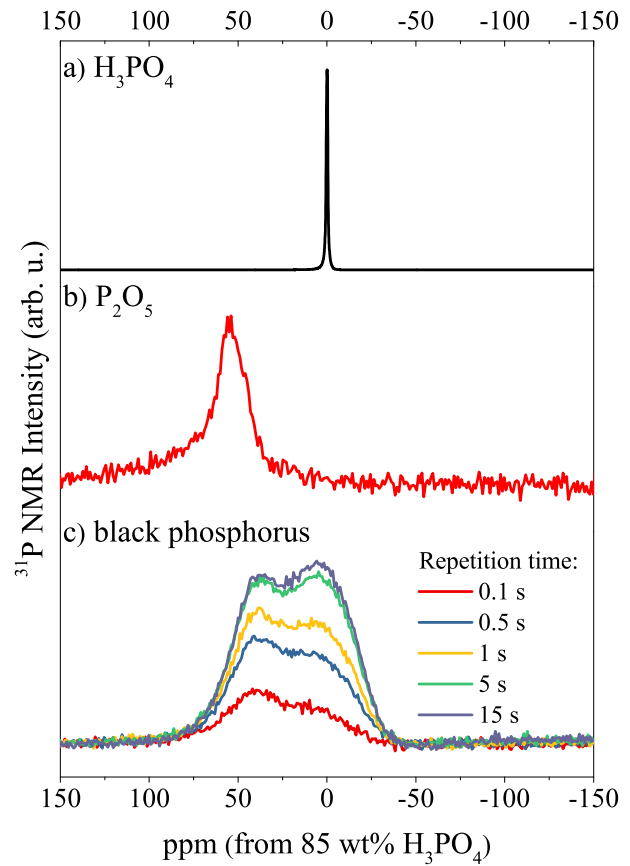}
\caption{$^{31}$P NMR spectra of a) aqueous $85$ wt$\%$ H$_3$PO$_4$ phosphoric acid, reference material for phosphor nuclei, b) dried P$_2$O$_5$ diphosphorus pentoxide, precursor of phosphoric acid, c) black phosphorus with different repetition times. The acid and the pentoxide are the most probable degradation products of bP under ambient conditions. The peaks in bP are located at $38(3)$ ppm and $4(3)$ ppm, while the oxide peak appears at $51(3)$ ppm relative to H$_3$PO$_4$. According to the spectra the bP sample is completely intact, no traces of H$_3$PO$_4$ or P$_2$O$_5$ is present (peaks are positioned at different NMR shifts). Our results are consistent with the current literature \cite{wang2016_1,andrew1966}. In the black phosphorus a doublet structure is observed, identified as a Pake doublet. The change of spectra upon different repetition times yield that the $T_1$ relaxation time is anisotropic.}
\label{fig:nmrspectra}
\end{figure}

Static $^{31}$P NMR spectra of H$_3$PO$_4$, P$_2$O$_5$ and bP are presented in Fig. \ref{fig:nmrspectra}. The phosphoric acid used as reference is positioned at $0$ ppm, the NMR shift of the other two materials are relative to this value. Black phosphorus exhibits a characteristic Pake doublet structure, which consist of two peaks: one at $38(3)$ ppm and one at $4(3)$ ppm. This result is consistent with the static NMR results of previous work done by Wang \emph{et al}. \cite{wang2016_1}. The features from bP are well distinguishable from the one found in P$_2$O$_5$, which is located at $51(3)$ ppm (in a good agreement with the literature value of $48$ ppm \cite{andrew1966}) and from the one in H$_3$PO$_4$. Phosphorus acid presents two features, one at about $10$ ppm and a second one at $5.5$ ppm \cite{wang2016_1}, which are also well distinguishable from the peaks of bP.

The $T_1$ spin-lattice relaxation time is acquired from free induction decay amplitudes with different pulse repetition times. The amplitude was obtained from Lorentzian fits of the peaks. Fig. \ref{fig:nmrspectra} demonstrates that the relaxation time is slightly non-uniform among the peaks of the doublet: $T_1(40~\text{ppm})=0.29(5)$ s and $T_1(6~\text{ppm})=0.7(1)$ s. The relaxation time for P$_2$O$_5$ is found to be $13(4)$ seconds, which is at least one magnitude longer. The shorter $T_1$ in bP as compared to P$_2$O$_5$ could be caused by a small amount of charge carriers in bP (via thermal excitation or defect doping) \cite{slichter}, while in the pentoxide it is not expected as oxides usually form large gap insulators. For the H$_3$PO$_4$ we found $0.8(2)$ s, which means that it is more diluted with water than its nominal value of $85$ wt$\%$ according to \cite{morgan1975}, fortunately the NMR shift is not affected by the concentration. The spin-spin relaxation time, $T_2$ is found to be homogeneous, within the error of the measurement in black phosphorus with a value about $50(7)$ {\textmu}s. Such a short $T_2$ is typical in solids with strong dipole-dipole coupling.

\subsection{Simulation of the doublet structure}

In solids, where dipole-dipole coupling interactions between like-nuclei (meaning: the same) is large, a line-broadening is observed due to the so-called van Vleck formula \cite{slichter}. The van Vleck formula is the result of several like-nuclei being distributed around a central nucleus. However, when only two nuclei form a strongly dipole coupled pair, the other nuclei being further away (as dipole field decays with $1/r^3$), a characteristic doublet structure is formed, the so-called Pake doublet.

\begin{figure}[h!]
\includegraphics*[width=.9\linewidth]{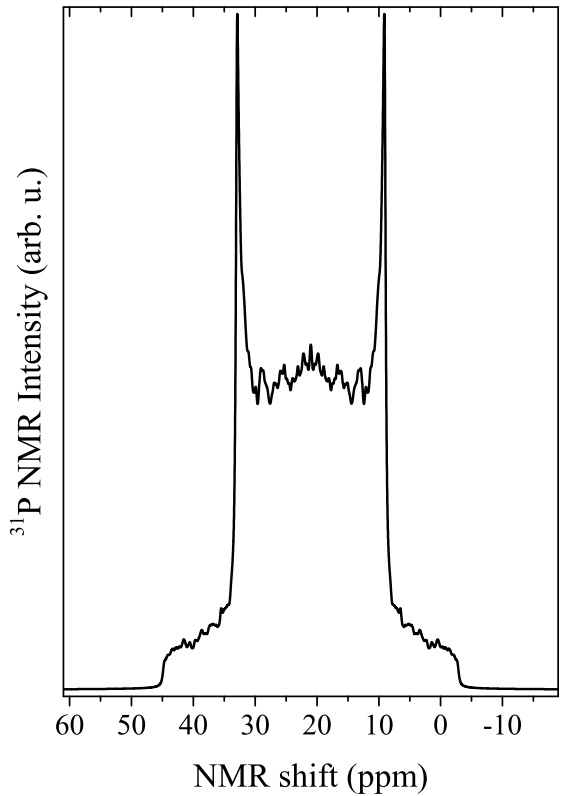}
\caption{Calculated NMR spectrum for a pair of $^{31}$P nuclei showing the characteristic Pake doublet structure arising from dipole-dipole interaction.}
\label{fig:sim}
\end{figure}

Fig. \ref{fig:sim} shows the calculated NMR spectrum for a pair of phosphorus nuclei (lattice constants taken from \cite{Castellanos-Gomez2014}). Note the characteristic doublet structure which reproduces qualitatively the experimental spectrum presented previously. A calculation of the NMR spectrum which includes several $^{31}$P nuclei is beyond the scope of the present contribution.

\subsection{ESR}

We recorded ESR signal from $50$ mT to $650$ mT of the black phosphorus in $X$-band (at $9$ GHz) with a high sensitivity spectrometer with various spectrometer configurations. Similar studies were successfully applied for the study of compounds with weak magnetism \cite{simon2006,simon2007,quintavalle2009,fabian2012,markus2015}. However, beside all efforts we did not record any ESR signal in bP not even that of impurity spins (e.g. from dangling bonds) which often occur in nanocarbon materials. On the other hand this also means that no conducting electrons are present, which is not so surprising taking into account that bP is a diamagnetic \cite{bridgman1914} semiconductor \cite{Keyes1953,Warschauer1963}. Another possibility is that any spin components, either localized or itinerant, could have a large ESR linewidth, which prevents their observations. We note that P is a relatively heavy element as compared to the lighter carbon, the spin-orbit couping, which usually affects ESR linewidths, could be much larger. During the experiments a drop in the $Q$-factor of the microwave cavity is also noticed from $Q_0=5000$ to $Q_{\text{L}}=600$, this is a sign of high microwave absorption in the sample, which implies that bP might find applications as a microwave absorbent.

\subsection{Conductivity}

\begin{figure}[h!]
\includegraphics*[width=.9\linewidth]{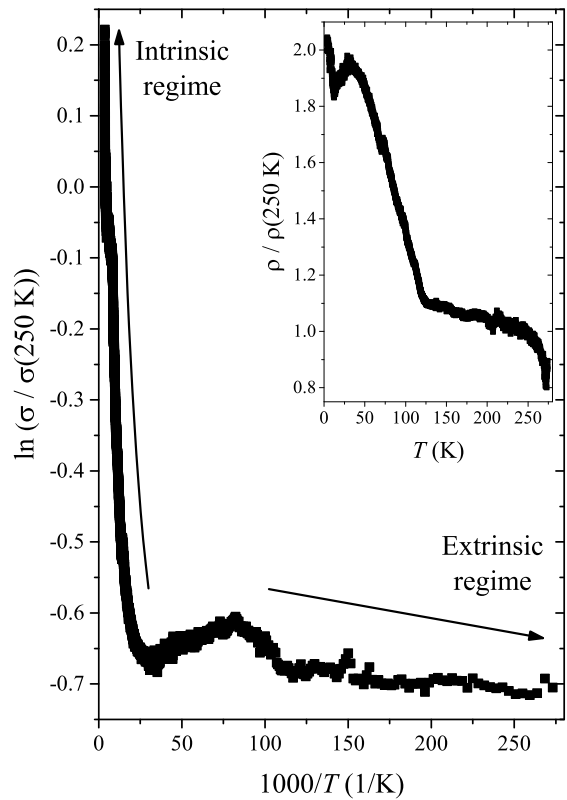}
\caption{Arrhenius type plot of conductivity results, normalized to the value at obtained $250$ K. The conductivity data exhibit a weakly doped semiconducting characteristics, thus two regimes can be identified with two different activation energies accordingly. The high temperature, intrinsic domain is arising from the band structure of black phosphorus, and has a gap of $0.36(1)$ eV in agreement to literature results on polycrystalline and single crystal materials \cite{Keyes1953,Warschauer1963,Maruyama1981,Akahama1983,Narita1983,Hayashi1984,Maruyama1986}. The extrinsic part at low temperatures is dominated by the activated conductance of charged impurities/defects. The size of this gap is $5.6(7)$ meV. The inset presents the resistance as a function of temperature.}
\label{fig:mwcond}
\end{figure}

Microwave conductivity of black phosphorus was measured in the range of $3.7-274$ K. The observed values were corrected with the conductivity coming from the empty copper cavity in the whole temperature range and was normalized to the resistivity value in bP at $250$ K. Then the data was transformed to an Arrhenius plot, which is shown in Fig. \ref{fig:mwcond}. The resistivity as a function of the temperature is plotted in the inset of Fig. \ref{fig:mwcond}. 

Black phosphorus exhibits a weakly doped semiconducting behavior in the whole examined temperature range, the resistivity drops with increasing temperature. At high temperature, above $120$ K the resistivity curve follows an activated behavior: $\rho=\rho_0 \exp{\left(E_{\text{g,i}}/2 k_{\text{B}}T\right)}$, where $E_{\text{g,i}}$ is the band gap and $k_{\text{B}}$ is the Boltzmann-constant. The value for $E_{\text{g,i}} = 0.36(1)$ eV is obtained through fitting the curve. This value agrees well with the literature: $0.33$ eV \cite{Keyes1953,Akahama1983,Hayashi1984} and $0.35$ eV \cite{Warschauer1963} found in polycrystalline materials, $0.31$ eV \cite{Maruyama1981}, $0.335$ eV \cite{Narita1983}, $0.34$ eV \cite{Maruyama1986} and $0.28$ eV \cite{Baba1989_1} found in single crystals, therefore this regime is identified as the intrinsic one.

Below $120$ K, the low temperature part also goes in a similar manner, except that the gap is $2$ orders of magnitude smaller: $E_{\text{g,e}} = 5.6(7)$ meV. This small value of the band gap indicates the presence of shallow donor or acceptor bands. This regime is thus identified as the extrinsic one. The possible dopants in our case are charged impurities, most probably coming from surface bound oxygen. According to Ziletti \emph{et al}. \cite{ziletti2015} a horizontal $-$P$-$O$-$P$-$ oxygen bridge can result a new donor type band near the conduction band. On the other hand Narita \emph{et al}. \cite{Narita1983}, Akahama \emph{et al.} \cite{Akahama1983} and Baba \emph{et al.} \cite{Baba1991_1} claimed that according to their Hall and conductivity measurements that the small gap is arising from hole dopants and the gap for the acceptor band is about $15-19.5$ meV, $11.1-18.9$ meV and $21.4(1)$, respectively, which agrees roughly with our results.

\section{Conclusions}

We presented NMR, ESR and microwave conductivity measurements on balck phosphorus. $^{31}$P NMR indicates a semiconducting/insulating behavior with the well-known Pake doublet structure which is characteristic for $I=1/2$ nuclei and also carries information about the local structure. The two features of the doublet present in the NMR spectrum are well distinguishable from the degradation products, namely from the oxides and acids. Spin-lattice and spin-spin relaxation times were also extracted. $T_1$ shows non-uniform behavior and also differs gradually from the other investigated species. ESR indicates the absence of free charge carriers and shows a negligible amount of impurities only and yields that the material is a good microwave absorber. The temperature dependent microwave conductivity showed that the material is semiconducting with an intrinsic gap of $0.36(1)$ eV and an extrinsic one with $5.6(7)$ meV. This double activated behavior is well known for weakly doped semiconductors.

\section{Acknowledgement}
Work supported by the Hungarian National Research, Development and Innovation Office (NKFIH) Grant Nr. K119442 and Nr. K107228. The authors thank the European Research Council (ERC Advanced Grant 742145 B-PhosphoChem) for financial support. The research leading to these results was partially funded by the European Union Seventh Framework Programme under grant agreement No. 604391 Graphene Flagship. We also thank the Deutsche Forschungsgemeinschaft (DFG-SFB 953 "Synthetic Carbon Allotropes", Project A1), the Interdisciplinary Center for Molecular Materials (ICMM), and the Graduate School Molecular Science (GSMS) for financial support. J. C. C.-T. thanks S. R. for the use of equipment, and acknowledge the financial support of the DRS Postdoc Fellowship Point-2014 of the NanoScale Focus Area at Freie Universit\"{a}t Berlin.

\bibliography{bp}

\begin{thebibliography}{58}%
\makeatletter
\providecommand \@ifxundefined [1]{%
 \@ifx{#1\undefined}
}%
\providecommand \@ifnum [1]{%
 \ifnum #1\expandafter \@firstoftwo
 \else \expandafter \@secondoftwo
 \fi
}%
\providecommand \@ifx [1]{%
 \ifx #1\expandafter \@firstoftwo
 \else \expandafter \@secondoftwo
 \fi
}%
\providecommand \natexlab [1]{#1}%
\providecommand \enquote  [1]{``#1''}%
\providecommand \bibnamefont  [1]{#1}%
\providecommand \bibfnamefont [1]{#1}%
\providecommand \citenamefont [1]{#1}%
\providecommand \href@noop [0]{\@secondoftwo}%
\providecommand \href [0]{\begingroup \@sanitize@url \@href}%
\providecommand \@href[1]{\@@startlink{#1}\@@href}%
\providecommand \@@href[1]{\endgroup#1\@@endlink}%
\providecommand \@sanitize@url [0]{\catcode `\\12\catcode `\$12\catcode
  `\&12\catcode `\#12\catcode `\^12\catcode `\_12\catcode `\%12\relax}%
\providecommand \@@startlink[1]{}%
\providecommand \@@endlink[0]{}%
\providecommand \url  [0]{\begingroup\@sanitize@url \@url }%
\providecommand \@url [1]{\endgroup\@href {#1}{\urlprefix }}%
\providecommand \urlprefix  [0]{URL }%
\providecommand \Eprint [0]{\href }%
\providecommand \doibase [0]{http://dx.doi.org/}%
\providecommand \selectlanguage [0]{\@gobble}%
\providecommand \bibinfo  [0]{\@secondoftwo}%
\providecommand \bibfield  [0]{\@secondoftwo}%
\providecommand \translation [1]{[#1]}%
\providecommand \BibitemOpen [0]{}%
\providecommand \bibitemStop [0]{}%
\providecommand \bibitemNoStop [0]{.\EOS\space}%
\providecommand \EOS [0]{\spacefactor3000\relax}%
\providecommand \BibitemShut  [1]{\csname bibitem#1\endcsname}%
\let\auto@bib@innerbib\@empty
\bibitem [{\citenamefont {Novoselov}\ \emph {et~al.}(2004)\citenamefont
  {Novoselov}, \citenamefont {Geim}, \citenamefont {Morozov}, \citenamefont
  {Jiang}, \citenamefont {Zhang}, \citenamefont {Dubonos}, \citenamefont
  {Grigorieva},\ and\ \citenamefont {Firsov}}]{novoselov2004}%
  \BibitemOpen
  \bibfield  {author} {\bibinfo {author} {\bibfnamefont {K.~S.}\ \bibnamefont
  {Novoselov}}, \bibinfo {author} {\bibfnamefont {A.~K.}\ \bibnamefont {Geim}},
  \bibinfo {author} {\bibfnamefont {S.~V.}\ \bibnamefont {Morozov}}, \bibinfo
  {author} {\bibfnamefont {D.}~\bibnamefont {Jiang}}, \bibinfo {author}
  {\bibfnamefont {Y.}~\bibnamefont {Zhang}}, \bibinfo {author} {\bibfnamefont
  {S.~V.}\ \bibnamefont {Dubonos}}, \bibinfo {author} {\bibfnamefont {I.~V.}\
  \bibnamefont {Grigorieva}}, \ and\ \bibinfo {author} {\bibfnamefont {A.~A.}\
  \bibnamefont {Firsov}},\ }\href@noop {} {\bibfield  {journal} {\bibinfo
  {journal} {Science}\ }\textbf {\bibinfo {volume} {306}},\ \bibinfo {pages}
  {666} (\bibinfo {year} {2004})}\BibitemShut {NoStop}%
\bibitem [{\citenamefont {Bridgman}(1914)}]{bridgman1914}%
  \BibitemOpen
  \bibfield  {author} {\bibinfo {author} {\bibfnamefont {P.~W.}\ \bibnamefont
  {Bridgman}},\ }\href@noop {} {\bibfield  {journal} {\bibinfo  {journal} {J.
  Am. Chem. Soc.}\ }\textbf {\bibinfo {volume} {36}},\ \bibinfo {pages} {1344}
  (\bibinfo {year} {1914})}\BibitemShut {NoStop}%
\bibitem [{\citenamefont {Liu}\ \emph {et~al.}(2014)\citenamefont {Liu},
  \citenamefont {Neal}, \citenamefont {Zhu}, \citenamefont {Luo}, \citenamefont
  {Xu}, \citenamefont {Tom\'{a}nek},\ and\ \citenamefont {Ye}}]{Liu2014}%
  \BibitemOpen
  \bibfield  {author} {\bibinfo {author} {\bibfnamefont {H.}~\bibnamefont
  {Liu}}, \bibinfo {author} {\bibfnamefont {A.~T.}\ \bibnamefont {Neal}},
  \bibinfo {author} {\bibfnamefont {Z.}~\bibnamefont {Zhu}}, \bibinfo {author}
  {\bibfnamefont {Z.}~\bibnamefont {Luo}}, \bibinfo {author} {\bibfnamefont
  {X.}~\bibnamefont {Xu}}, \bibinfo {author} {\bibfnamefont {D.}~\bibnamefont
  {Tom\'{a}nek}}, \ and\ \bibinfo {author} {\bibfnamefont {P.~D.}\ \bibnamefont
  {Ye}},\ }\href@noop {} {\bibfield  {journal} {\bibinfo  {journal} {ACS Nano}\
  }\textbf {\bibinfo {volume} {8}},\ \bibinfo {pages} {4033} (\bibinfo {year}
  {2014})}\BibitemShut {NoStop}%
\bibitem [{\citenamefont {Ling}\ \emph {et~al.}(2015)\citenamefont {Ling},
  \citenamefont {Wang}, \citenamefont {Huang}, \citenamefont {Xia},\ and\
  \citenamefont {Dresselhaus}}]{ling2015}%
  \BibitemOpen
  \bibfield  {author} {\bibinfo {author} {\bibfnamefont {X.}~\bibnamefont
  {Ling}}, \bibinfo {author} {\bibfnamefont {H.}~\bibnamefont {Wang}}, \bibinfo
  {author} {\bibfnamefont {S.}~\bibnamefont {Huang}}, \bibinfo {author}
  {\bibfnamefont {F.}~\bibnamefont {Xia}}, \ and\ \bibinfo {author}
  {\bibfnamefont {M.~S.}\ \bibnamefont {Dresselhaus}},\ }\href@noop {}
  {\bibfield  {journal} {\bibinfo  {journal} {Proc. Nat. Acad. Sci. USA
  (PNAS)}\ }\textbf {\bibinfo {volume} {112}},\ \bibinfo {pages} {4523}
  (\bibinfo {year} {2015})}\BibitemShut {NoStop}%
\bibitem [{\citenamefont {Gusm$\tilde{\text{a}}$o}\ \emph
  {et~al.}(2017)\citenamefont {Gusm$\tilde{\text{a}}$o}, \citenamefont
  {Sofer},\ and\ \citenamefont {Pumera}}]{gusmao2017}%
  \BibitemOpen
  \bibfield  {author} {\bibinfo {author} {\bibfnamefont {R.}~\bibnamefont
  {Gusm$\tilde{\text{a}}$o}}, \bibinfo {author} {\bibfnamefont
  {Z.}~\bibnamefont {Sofer}}, \ and\ \bibinfo {author} {\bibfnamefont
  {M.}~\bibnamefont {Pumera}},\ }\href@noop {} {\bibfield  {journal} {\bibinfo
  {journal} {Angew. Chem. Int. Ed.}\ ,\ \bibinfo {pages} {201610512}} (\bibinfo
  {year} {2017})}\BibitemShut {NoStop}%
\bibitem [{\citenamefont {Castellanos-Gomez}(2015)}]{castellanos-gomez2015}%
  \BibitemOpen
  \bibfield  {author} {\bibinfo {author} {\bibfnamefont {A.}~\bibnamefont
  {Castellanos-Gomez}},\ }\href@noop {} {\bibfield  {journal} {\bibinfo
  {journal} {J. Phys. Chem. Lett.}\ }\textbf {\bibinfo {volume} {6}},\ \bibinfo
  {pages} {4280} (\bibinfo {year} {2015})}\BibitemShut {NoStop}%
\bibitem [{\citenamefont {Li}\ \emph {et~al.}(2014)\citenamefont {Li},
  \citenamefont {Yu}, \citenamefont {Ye}, \citenamefont {Ge}, \citenamefont
  {Ou}, \citenamefont {Wu}, \citenamefont {Feng}, \citenamefont {Chen},\ and\
  \citenamefont {Zhang}}]{Li2014}%
  \BibitemOpen
  \bibfield  {author} {\bibinfo {author} {\bibfnamefont {L.}~\bibnamefont
  {Li}}, \bibinfo {author} {\bibfnamefont {Y.}~\bibnamefont {Yu}}, \bibinfo
  {author} {\bibfnamefont {G.~J.}\ \bibnamefont {Ye}}, \bibinfo {author}
  {\bibfnamefont {Q.}~\bibnamefont {Ge}}, \bibinfo {author} {\bibfnamefont
  {X.}~\bibnamefont {Ou}}, \bibinfo {author} {\bibfnamefont {H.}~\bibnamefont
  {Wu}}, \bibinfo {author} {\bibfnamefont {D.}~\bibnamefont {Feng}}, \bibinfo
  {author} {\bibfnamefont {X.~H.}\ \bibnamefont {Chen}}, \ and\ \bibinfo
  {author} {\bibfnamefont {Y.}~\bibnamefont {Zhang}},\ }\href@noop {}
  {\bibfield  {journal} {\bibinfo  {journal} {Nat. Nanotechnol.}\ }\textbf
  {\bibinfo {volume} {9}},\ \bibinfo {pages} {372} (\bibinfo {year}
  {2014})}\BibitemShut {NoStop}%
\bibitem [{\citenamefont {Lu}\ \emph {et~al.}(2014)\citenamefont {Lu},
  \citenamefont {Nan}, \citenamefont {Hong}, \citenamefont {Chen},
  \citenamefont {Zhu}, \citenamefont {Liang}, \citenamefont {Ma}, \citenamefont
  {Ni}, \citenamefont {Jin},\ and\ \citenamefont {Zhang}}]{Lu2014}%
  \BibitemOpen
  \bibfield  {author} {\bibinfo {author} {\bibfnamefont {W.}~\bibnamefont
  {Lu}}, \bibinfo {author} {\bibfnamefont {H.}~\bibnamefont {Nan}}, \bibinfo
  {author} {\bibfnamefont {J.}~\bibnamefont {Hong}}, \bibinfo {author}
  {\bibfnamefont {Y.}~\bibnamefont {Chen}}, \bibinfo {author} {\bibfnamefont
  {C.}~\bibnamefont {Zhu}}, \bibinfo {author} {\bibfnamefont {Z.}~\bibnamefont
  {Liang}}, \bibinfo {author} {\bibfnamefont {X.}~\bibnamefont {Ma}}, \bibinfo
  {author} {\bibfnamefont {Z.}~\bibnamefont {Ni}}, \bibinfo {author}
  {\bibfnamefont {C.}~\bibnamefont {Jin}}, \ and\ \bibinfo {author}
  {\bibfnamefont {Z.}~\bibnamefont {Zhang}},\ }\href@noop {} {\bibfield
  {journal} {\bibinfo  {journal} {Nano Res.}\ }\textbf {\bibinfo {volume}
  {7}},\ \bibinfo {pages} {853} (\bibinfo {year} {2014})}\BibitemShut {NoStop}%
\bibitem [{\citenamefont {Castellanos-Gomez}\ \emph {et~al.}(2014)\citenamefont
  {Castellanos-Gomez}, \citenamefont {Vicarelli}, \citenamefont {Prada},
  \citenamefont {Island}, \citenamefont {Narasimha-Acharya}, \citenamefont
  {Blanter}, \citenamefont {Groenendijk}, \citenamefont {Buscema},
  \citenamefont {Steele}, \citenamefont {Alvarez}, \citenamefont {Zandbergen},
  \citenamefont {Palacios},\ and\ \citenamefont
  {{{van~der~Zant}}}}]{Castellanos-Gomez2014}%
  \BibitemOpen
  \bibfield  {author} {\bibinfo {author} {\bibfnamefont {A.}~\bibnamefont
  {Castellanos-Gomez}}, \bibinfo {author} {\bibfnamefont {L.}~\bibnamefont
  {Vicarelli}}, \bibinfo {author} {\bibfnamefont {E.}~\bibnamefont {Prada}},
  \bibinfo {author} {\bibfnamefont {J.~O.}\ \bibnamefont {Island}}, \bibinfo
  {author} {\bibfnamefont {K.~L.}\ \bibnamefont {Narasimha-Acharya}}, \bibinfo
  {author} {\bibfnamefont {S.~I.}\ \bibnamefont {Blanter}}, \bibinfo {author}
  {\bibfnamefont {D.~J.}\ \bibnamefont {Groenendijk}}, \bibinfo {author}
  {\bibfnamefont {M.}~\bibnamefont {Buscema}}, \bibinfo {author} {\bibfnamefont
  {G.~A.}\ \bibnamefont {Steele}}, \bibinfo {author} {\bibfnamefont {J.~V.}\
  \bibnamefont {Alvarez}}, \bibinfo {author} {\bibfnamefont {H.~W.}\
  \bibnamefont {Zandbergen}}, \bibinfo {author} {\bibfnamefont {J.~J.}\
  \bibnamefont {Palacios}}, \ and\ \bibinfo {author} {\bibfnamefont {H.~S.~J.}\
  \bibnamefont {{{van~der~Zant}}}},\ }\href@noop {} {\bibfield  {journal}
  {\bibinfo  {journal} {2D Mater.}\ }\textbf {\bibinfo {volume} {1}},\ \bibinfo
  {pages} {025001} (\bibinfo {year} {2014})}\BibitemShut {NoStop}%
\bibitem [{\citenamefont {Koenig}\ \emph {et~al.}(2014)\citenamefont {Koenig},
  \citenamefont {Doganov}, \citenamefont {Schmidt}, \citenamefont
  {{{Castro~Neto}}},\ and\ \citenamefont {\"{O}zyilmaz}}]{Koening2014}%
  \BibitemOpen
  \bibfield  {author} {\bibinfo {author} {\bibfnamefont {S.~P.}\ \bibnamefont
  {Koenig}}, \bibinfo {author} {\bibfnamefont {R.~A.}\ \bibnamefont {Doganov}},
  \bibinfo {author} {\bibfnamefont {H.}~\bibnamefont {Schmidt}}, \bibinfo
  {author} {\bibfnamefont {A.~H.}\ \bibnamefont {{{Castro~Neto}}}}, \ and\
  \bibinfo {author} {\bibfnamefont {B.}~\bibnamefont {\"{O}zyilmaz}},\
  }\href@noop {} {\bibfield  {journal} {\bibinfo  {journal} {Appl. Phys.
  Lett.}\ }\textbf {\bibinfo {volume} {104}},\ \bibinfo {pages} {103106}
  (\bibinfo {year} {2014})}\BibitemShut {NoStop}%
\bibitem [{\citenamefont {Buscema}\ \emph {et~al.}(2014)\citenamefont
  {Buscema}, \citenamefont {Groenendijk}, \citenamefont {Blanter},
  \citenamefont {Steele}, \citenamefont {{{van~der~Zant}}},\ and\ \citenamefont
  {Castellanos-Gomez}}]{Buscema2014}%
  \BibitemOpen
  \bibfield  {author} {\bibinfo {author} {\bibfnamefont {M.}~\bibnamefont
  {Buscema}}, \bibinfo {author} {\bibfnamefont {D.~J.}\ \bibnamefont
  {Groenendijk}}, \bibinfo {author} {\bibfnamefont {S.~I.}\ \bibnamefont
  {Blanter}}, \bibinfo {author} {\bibfnamefont {G.~A.}\ \bibnamefont {Steele}},
  \bibinfo {author} {\bibfnamefont {H.~S.~J.}\ \bibnamefont
  {{{van~der~Zant}}}}, \ and\ \bibinfo {author} {\bibfnamefont
  {A.}~\bibnamefont {Castellanos-Gomez}},\ }\href@noop {} {\bibfield  {journal}
  {\bibinfo  {journal} {Nano Lett.}\ }\textbf {\bibinfo {volume} {14}},\
  \bibinfo {pages} {3347} (\bibinfo {year} {2014})}\BibitemShut {NoStop}%
\bibitem [{\citenamefont {Zhang}\ \emph {et~al.}(2014)\citenamefont {Zhang},
  \citenamefont {Yang}, \citenamefont {Xu}, \citenamefont {Wang}, \citenamefont
  {Li}, \citenamefont {Ghufran}, \citenamefont {Zhang}, \citenamefont {Yu},
  \citenamefont {Zhang}, \citenamefont {Qin},\ and\ \citenamefont
  {Lu}}]{zhang2014}%
  \BibitemOpen
  \bibfield  {author} {\bibinfo {author} {\bibfnamefont {S.}~\bibnamefont
  {Zhang}}, \bibinfo {author} {\bibfnamefont {J.}~\bibnamefont {Yang}},
  \bibinfo {author} {\bibfnamefont {R.}~\bibnamefont {Xu}}, \bibinfo {author}
  {\bibfnamefont {F.}~\bibnamefont {Wang}}, \bibinfo {author} {\bibfnamefont
  {W.}~\bibnamefont {Li}}, \bibinfo {author} {\bibfnamefont {M.}~\bibnamefont
  {Ghufran}}, \bibinfo {author} {\bibfnamefont {Y.-W.}\ \bibnamefont {Zhang}},
  \bibinfo {author} {\bibfnamefont {Z.}~\bibnamefont {Yu}}, \bibinfo {author}
  {\bibfnamefont {G.}~\bibnamefont {Zhang}}, \bibinfo {author} {\bibfnamefont
  {Q.}~\bibnamefont {Qin}}, \ and\ \bibinfo {author} {\bibfnamefont
  {Y.}~\bibnamefont {Lu}},\ }\href@noop {} {\bibfield  {journal} {\bibinfo
  {journal} {ACS Nano}\ }\textbf {\bibinfo {volume} {8}},\ \bibinfo {pages}
  {9590} (\bibinfo {year} {2014})}\BibitemShut {NoStop}%
\bibitem [{\citenamefont {Brent}\ \emph {et~al.}(2014)\citenamefont {Brent},
  \citenamefont {Savjani}, \citenamefont {Lewis}, \citenamefont {Haigh},
  \citenamefont {Lewis},\ and\ \citenamefont {O'Brien}}]{brent2014}%
  \BibitemOpen
  \bibfield  {author} {\bibinfo {author} {\bibfnamefont {J.~R.}\ \bibnamefont
  {Brent}}, \bibinfo {author} {\bibfnamefont {N.}~\bibnamefont {Savjani}},
  \bibinfo {author} {\bibfnamefont {E.~A.}\ \bibnamefont {Lewis}}, \bibinfo
  {author} {\bibfnamefont {S.~J.}\ \bibnamefont {Haigh}}, \bibinfo {author}
  {\bibfnamefont {D.~J.}\ \bibnamefont {Lewis}}, \ and\ \bibinfo {author}
  {\bibfnamefont {P.}~\bibnamefont {O'Brien}},\ }\href@noop {} {\bibfield
  {journal} {\bibinfo  {journal} {Chem. Commun.}\ }\textbf {\bibinfo {volume}
  {50}},\ \bibinfo {pages} {13338} (\bibinfo {year} {2014})}\BibitemShut
  {NoStop}%
\bibitem [{\citenamefont {Yasaei}\ \emph {et~al.}(2015)\citenamefont {Yasaei},
  \citenamefont {Kumar}, \citenamefont {Foroozan}, \citenamefont {Wang},
  \citenamefont {Asadi}, \citenamefont {Tuschel}, \citenamefont {Indacochea},
  \citenamefont {Klie},\ and\ \citenamefont {Salehi-Khojin}}]{yasaei2015}%
  \BibitemOpen
  \bibfield  {author} {\bibinfo {author} {\bibfnamefont {P.}~\bibnamefont
  {Yasaei}}, \bibinfo {author} {\bibfnamefont {B.}~\bibnamefont {Kumar}},
  \bibinfo {author} {\bibfnamefont {T.}~\bibnamefont {Foroozan}}, \bibinfo
  {author} {\bibfnamefont {C.}~\bibnamefont {Wang}}, \bibinfo {author}
  {\bibfnamefont {M.}~\bibnamefont {Asadi}}, \bibinfo {author} {\bibfnamefont
  {D.}~\bibnamefont {Tuschel}}, \bibinfo {author} {\bibfnamefont {J.~E.}\
  \bibnamefont {Indacochea}}, \bibinfo {author} {\bibfnamefont {R.~F.}\
  \bibnamefont {Klie}}, \ and\ \bibinfo {author} {\bibfnamefont
  {A.}~\bibnamefont {Salehi-Khojin}},\ }\href@noop {} {\bibfield  {journal}
  {\bibinfo  {journal} {Adv. Mater.}\ }\textbf {\bibinfo {volume} {27}},\
  \bibinfo {pages} {1887} (\bibinfo {year} {2015})}\BibitemShut {NoStop}%
\bibitem [{\citenamefont {Kang}\ \emph {et~al.}(2015)\citenamefont {Kang},
  \citenamefont {D.Wood}, \citenamefont {Wells}, \citenamefont {Lee},
  \citenamefont {Liu}, \citenamefont {Chen},\ and\ \citenamefont
  {Hersam}}]{kang2015}%
  \BibitemOpen
  \bibfield  {author} {\bibinfo {author} {\bibfnamefont {J.}~\bibnamefont
  {Kang}}, \bibinfo {author} {\bibfnamefont {J.}~\bibnamefont {D.Wood}},
  \bibinfo {author} {\bibfnamefont {S.~A.}\ \bibnamefont {Wells}}, \bibinfo
  {author} {\bibfnamefont {J.-H.}\ \bibnamefont {Lee}}, \bibinfo {author}
  {\bibfnamefont {X.}~\bibnamefont {Liu}}, \bibinfo {author} {\bibfnamefont
  {K.-S.}\ \bibnamefont {Chen}}, \ and\ \bibinfo {author} {\bibfnamefont
  {M.~C.}\ \bibnamefont {Hersam}},\ }\href@noop {} {\bibfield  {journal}
  {\bibinfo  {journal} {ACS Nano}\ }\textbf {\bibinfo {volume} {9}},\ \bibinfo
  {pages} {3596} (\bibinfo {year} {2015})}\BibitemShut {NoStop}%
\bibitem [{\citenamefont {Hanlon}\ \emph {et~al.}(2015)\citenamefont {Hanlon},
  \citenamefont {Backes}, \citenamefont {Doherty}, \citenamefont {Cucinotta},
  \citenamefont {Berner}, \citenamefont {Boland}, \citenamefont {Lee},
  \citenamefont {Harvey}, \citenamefont {Lynch}, \citenamefont {Gholamvand},
  \citenamefont {Zhang}, \citenamefont {Wang}, \citenamefont {Moynihan},
  \citenamefont {Pokle}, \citenamefont {Ramasse}, \citenamefont {McEvoy},
  \citenamefont {Blau}, \citenamefont {Wang}, \citenamefont {Abellan},
  \citenamefont {Hauke}, \citenamefont {Hirsch}, \citenamefont {Sanvito},
  \citenamefont {O'Regan}, \citenamefont {Duesberg}, \citenamefont {Nicolosi},\
  and\ \citenamefont {Coleman}}]{hanlon2015}%
  \BibitemOpen
  \bibfield  {author} {\bibinfo {author} {\bibfnamefont {D.}~\bibnamefont
  {Hanlon}}, \bibinfo {author} {\bibfnamefont {C.}~\bibnamefont {Backes}},
  \bibinfo {author} {\bibfnamefont {E.}~\bibnamefont {Doherty}}, \bibinfo
  {author} {\bibfnamefont {C.~S.}\ \bibnamefont {Cucinotta}}, \bibinfo {author}
  {\bibfnamefont {N.~C.}\ \bibnamefont {Berner}}, \bibinfo {author}
  {\bibfnamefont {C.}~\bibnamefont {Boland}}, \bibinfo {author} {\bibfnamefont
  {K.}~\bibnamefont {Lee}}, \bibinfo {author} {\bibfnamefont {A.}~\bibnamefont
  {Harvey}}, \bibinfo {author} {\bibfnamefont {P.}~\bibnamefont {Lynch}},
  \bibinfo {author} {\bibfnamefont {Z.}~\bibnamefont {Gholamvand}}, \bibinfo
  {author} {\bibfnamefont {S.}~\bibnamefont {Zhang}}, \bibinfo {author}
  {\bibfnamefont {K.}~\bibnamefont {Wang}}, \bibinfo {author} {\bibfnamefont
  {G.}~\bibnamefont {Moynihan}}, \bibinfo {author} {\bibfnamefont
  {A.}~\bibnamefont {Pokle}}, \bibinfo {author} {\bibfnamefont {Q.~M.}\
  \bibnamefont {Ramasse}}, \bibinfo {author} {\bibfnamefont {N.}~\bibnamefont
  {McEvoy}}, \bibinfo {author} {\bibfnamefont {W.~J.}\ \bibnamefont {Blau}},
  \bibinfo {author} {\bibfnamefont {J.}~\bibnamefont {Wang}}, \bibinfo {author}
  {\bibfnamefont {G.}~\bibnamefont {Abellan}}, \bibinfo {author} {\bibfnamefont
  {F.}~\bibnamefont {Hauke}}, \bibinfo {author} {\bibfnamefont
  {A.}~\bibnamefont {Hirsch}}, \bibinfo {author} {\bibfnamefont
  {S.}~\bibnamefont {Sanvito}}, \bibinfo {author} {\bibfnamefont {D.~D.}\
  \bibnamefont {O'Regan}}, \bibinfo {author} {\bibfnamefont {G.~S.}\
  \bibnamefont {Duesberg}}, \bibinfo {author} {\bibfnamefont {V.}~\bibnamefont
  {Nicolosi}}, \ and\ \bibinfo {author} {\bibfnamefont {J.~N.}\ \bibnamefont
  {Coleman}},\ }\href@noop {} {\bibfield  {journal} {\bibinfo  {journal} {Nat.
  Commun.}\ }\textbf {\bibinfo {volume} {6}},\ \bibinfo {pages} {8563}
  (\bibinfo {year} {2015})}\BibitemShut {NoStop}%
\bibitem [{\citenamefont {Abell\'{a}n}\ \emph {et~al.}(2016)\citenamefont
  {Abell\'{a}n}, \citenamefont {Lloret}, \citenamefont {Mundloch},
  \citenamefont {Marcia}, \citenamefont {Neiss}, \citenamefont {G\"{o}rling},
  \citenamefont {Varela}, \citenamefont {Hauke},\ and\ \citenamefont
  {Hirsch}}]{Abellan2016}%
  \BibitemOpen
  \bibfield  {author} {\bibinfo {author} {\bibfnamefont {G.}~\bibnamefont
  {Abell\'{a}n}}, \bibinfo {author} {\bibfnamefont {V.}~\bibnamefont {Lloret}},
  \bibinfo {author} {\bibfnamefont {U.}~\bibnamefont {Mundloch}}, \bibinfo
  {author} {\bibfnamefont {M.}~\bibnamefont {Marcia}}, \bibinfo {author}
  {\bibfnamefont {C.}~\bibnamefont {Neiss}}, \bibinfo {author} {\bibfnamefont
  {A.}~\bibnamefont {G\"{o}rling}}, \bibinfo {author} {\bibfnamefont
  {M.}~\bibnamefont {Varela}}, \bibinfo {author} {\bibfnamefont
  {F.}~\bibnamefont {Hauke}}, \ and\ \bibinfo {author} {\bibfnamefont
  {A.}~\bibnamefont {Hirsch}},\ }\href@noop {} {\bibfield  {journal} {\bibinfo
  {journal} {Angew. Chem.}\ }\textbf {\bibinfo {volume} {128}},\ \bibinfo
  {pages} {1} (\bibinfo {year} {2016})}\BibitemShut {NoStop}%
\bibitem [{\citenamefont {Edmonds}\ \emph {et~al.}(2015)\citenamefont
  {Edmonds}, \citenamefont {Tadich}, \citenamefont {Carvalho}, \citenamefont
  {Ziletti}, \citenamefont {O'Donnell}, \citenamefont {Koenig}, \citenamefont
  {\"{O}zyilmaz}, \citenamefont {{{Castro~Neto}}},\ and\ \citenamefont
  {Fuhrer}}]{edmonds2015}%
  \BibitemOpen
  \bibfield  {author} {\bibinfo {author} {\bibfnamefont {M.~T.}\ \bibnamefont
  {Edmonds}}, \bibinfo {author} {\bibfnamefont {A.}~\bibnamefont {Tadich}},
  \bibinfo {author} {\bibfnamefont {A.}~\bibnamefont {Carvalho}}, \bibinfo
  {author} {\bibfnamefont {A.}~\bibnamefont {Ziletti}}, \bibinfo {author}
  {\bibfnamefont {K.~M.}\ \bibnamefont {O'Donnell}}, \bibinfo {author}
  {\bibfnamefont {S.~P.}\ \bibnamefont {Koenig}}, \bibinfo {author}
  {\bibfnamefont {D.~F. C.~B.}\ \bibnamefont {\"{O}zyilmaz}}, \bibinfo {author}
  {\bibfnamefont {A.~H.}\ \bibnamefont {{{Castro~Neto}}}}, \ and\ \bibinfo
  {author} {\bibfnamefont {M.~S.}\ \bibnamefont {Fuhrer}},\ }\href@noop {}
  {\bibfield  {journal} {\bibinfo  {journal} {ACS Appl. Mater. Interfaces}\
  }\textbf {\bibinfo {volume} {7}},\ \bibinfo {pages} {14557} (\bibinfo {year}
  {2015})}\BibitemShut {NoStop}%
\bibitem [{\citenamefont {Ziletti}\ \emph
  {et~al.}(2015{\natexlab{a}})\citenamefont {Ziletti}, \citenamefont
  {Carvalho}, \citenamefont {Campbell}, \citenamefont {Coker},\ and\
  \citenamefont {{{Castro~Neto}}}}]{ziletti2015}%
  \BibitemOpen
  \bibfield  {author} {\bibinfo {author} {\bibfnamefont {A.}~\bibnamefont
  {Ziletti}}, \bibinfo {author} {\bibfnamefont {A.}~\bibnamefont {Carvalho}},
  \bibinfo {author} {\bibfnamefont {D.~K.}\ \bibnamefont {Campbell}}, \bibinfo
  {author} {\bibfnamefont {D.~F.}\ \bibnamefont {Coker}}, \ and\ \bibinfo
  {author} {\bibfnamefont {A.~H.}\ \bibnamefont {{{Castro~Neto}}}},\
  }\href@noop {} {\bibfield  {journal} {\bibinfo  {journal} {Phys. Rev. Lett.}\
  }\textbf {\bibinfo {volume} {114}},\ \bibinfo {pages} {046801} (\bibinfo
  {year} {2015}{\natexlab{a}})}\BibitemShut {NoStop}%
\bibitem [{\citenamefont {Favron}\ \emph {et~al.}(2015)\citenamefont {Favron},
  \citenamefont {Gaufr\`{e}s}, \citenamefont {Fossard}, \citenamefont
  {Phaneuf-L'Heureux}, \citenamefont {Tang}, \citenamefont {L\'{e}vesque},
  \citenamefont {Loiseau}, \citenamefont {Leonelli}, \citenamefont
  {Francoeur},\ and\ \citenamefont {Martel}}]{favron2015}%
  \BibitemOpen
  \bibfield  {author} {\bibinfo {author} {\bibfnamefont {A.}~\bibnamefont
  {Favron}}, \bibinfo {author} {\bibfnamefont {E.}~\bibnamefont {Gaufr\`{e}s}},
  \bibinfo {author} {\bibfnamefont {F.}~\bibnamefont {Fossard}}, \bibinfo
  {author} {\bibfnamefont {A.-L.}\ \bibnamefont {Phaneuf-L'Heureux}}, \bibinfo
  {author} {\bibfnamefont {N.~Y.-W.}\ \bibnamefont {Tang}}, \bibinfo {author}
  {\bibfnamefont {P.~L.}\ \bibnamefont {L\'{e}vesque}}, \bibinfo {author}
  {\bibfnamefont {A.}~\bibnamefont {Loiseau}}, \bibinfo {author} {\bibfnamefont
  {R.}~\bibnamefont {Leonelli}}, \bibinfo {author} {\bibfnamefont
  {S.}~\bibnamefont {Francoeur}}, \ and\ \bibinfo {author} {\bibfnamefont
  {R.}~\bibnamefont {Martel}},\ }\href@noop {} {\bibfield  {journal} {\bibinfo
  {journal} {Nat. Mater.}\ }\textbf {\bibinfo {volume} {14}},\ \bibinfo {pages}
  {826} (\bibinfo {year} {2015})}\BibitemShut {NoStop}%
\bibitem [{\citenamefont {Wood}\ \emph {et~al.}(2014)\citenamefont {Wood},
  \citenamefont {Wells}, \citenamefont {Jariwala}, \citenamefont {Chen},
  \citenamefont {Cho}, \citenamefont {Sangwan}, \citenamefont {Liu},
  \citenamefont {Lauhon}, \citenamefont {Marks},\ and\ \citenamefont
  {Hersam}}]{wood2014}%
  \BibitemOpen
  \bibfield  {author} {\bibinfo {author} {\bibfnamefont {J.~D.}\ \bibnamefont
  {Wood}}, \bibinfo {author} {\bibfnamefont {S.~A.}\ \bibnamefont {Wells}},
  \bibinfo {author} {\bibfnamefont {D.}~\bibnamefont {Jariwala}}, \bibinfo
  {author} {\bibfnamefont {K.-S.}\ \bibnamefont {Chen}}, \bibinfo {author}
  {\bibfnamefont {E.}~\bibnamefont {Cho}}, \bibinfo {author} {\bibfnamefont
  {V.~K.}\ \bibnamefont {Sangwan}}, \bibinfo {author} {\bibfnamefont
  {X.}~\bibnamefont {Liu}}, \bibinfo {author} {\bibfnamefont {L.~J.}\
  \bibnamefont {Lauhon}}, \bibinfo {author} {\bibfnamefont {T.~J.}\
  \bibnamefont {Marks}}, \ and\ \bibinfo {author} {\bibfnamefont {M.~C.}\
  \bibnamefont {Hersam}},\ }\href@noop {} {\bibfield  {journal} {\bibinfo
  {journal} {Nano Lett.}\ }\textbf {\bibinfo {volume} {14}},\ \bibinfo {pages}
  {6964} (\bibinfo {year} {2014})}\BibitemShut {NoStop}%
\bibitem [{\citenamefont {Ziletti}\ \emph
  {et~al.}(2015{\natexlab{b}})\citenamefont {Ziletti}, \citenamefont
  {Carvalho}, \citenamefont {Trevisanutto}, \citenamefont {Campbell},
  \citenamefont {Coker},\ and\ \citenamefont
  {{{Castro~Neto}}}}]{ziletti2015_2}%
  \BibitemOpen
  \bibfield  {author} {\bibinfo {author} {\bibfnamefont {A.}~\bibnamefont
  {Ziletti}}, \bibinfo {author} {\bibfnamefont {A.}~\bibnamefont {Carvalho}},
  \bibinfo {author} {\bibfnamefont {P.~E.}\ \bibnamefont {Trevisanutto}},
  \bibinfo {author} {\bibfnamefont {D.~K.}\ \bibnamefont {Campbell}}, \bibinfo
  {author} {\bibfnamefont {D.~F.}\ \bibnamefont {Coker}}, \ and\ \bibinfo
  {author} {\bibfnamefont {A.~H.}\ \bibnamefont {{{Castro~Neto}}}},\
  }\href@noop {} {\bibfield  {journal} {\bibinfo  {journal} {Phys. Rev. B}\
  }\textbf {\bibinfo {volume} {91}},\ \bibinfo {pages} {085407} (\bibinfo
  {year} {2015}{\natexlab{b}})}\BibitemShut {NoStop}%
\bibitem [{\citenamefont {Wang}\ \emph
  {et~al.}(2016{\natexlab{a}})\citenamefont {Wang}, \citenamefont {Yang},
  \citenamefont {Wan}, \citenamefont {Xi}, \citenamefont {Zeng}, \citenamefont
  {Liu}, \citenamefont {Wu}, \citenamefont {Liu},\ and\ \citenamefont
  {Wang}}]{wang2016_1}%
  \BibitemOpen
  \bibfield  {author} {\bibinfo {author} {\bibfnamefont {Y.}~\bibnamefont
  {Wang}}, \bibinfo {author} {\bibfnamefont {B.}~\bibnamefont {Yang}}, \bibinfo
  {author} {\bibfnamefont {B.}~\bibnamefont {Wan}}, \bibinfo {author}
  {\bibfnamefont {X.}~\bibnamefont {Xi}}, \bibinfo {author} {\bibfnamefont
  {Z.}~\bibnamefont {Zeng}}, \bibinfo {author} {\bibfnamefont {E.}~\bibnamefont
  {Liu}}, \bibinfo {author} {\bibfnamefont {G.}~\bibnamefont {Wu}}, \bibinfo
  {author} {\bibfnamefont {Z.}~\bibnamefont {Liu}}, \ and\ \bibinfo {author}
  {\bibfnamefont {W.}~\bibnamefont {Wang}},\ }\href@noop {} {\bibfield
  {journal} {\bibinfo  {journal} {2D Mater.}\ }\textbf {\bibinfo {volume}
  {3}},\ \bibinfo {pages} {035025} (\bibinfo {year}
  {2016}{\natexlab{a}})}\BibitemShut {NoStop}%
\bibitem [{\citenamefont {Wang}\ \emph
  {et~al.}(2016{\natexlab{b}})\citenamefont {Wang}, \citenamefont {Slough},
  \citenamefont {Pandey},\ and\ \citenamefont {Karna}}]{wang2016_2}%
  \BibitemOpen
  \bibfield  {author} {\bibinfo {author} {\bibfnamefont {G.}~\bibnamefont
  {Wang}}, \bibinfo {author} {\bibfnamefont {W.~J.}\ \bibnamefont {Slough}},
  \bibinfo {author} {\bibfnamefont {R.}~\bibnamefont {Pandey}}, \ and\ \bibinfo
  {author} {\bibfnamefont {S.~P.}\ \bibnamefont {Karna}},\ }\href@noop {}
  {\bibfield  {journal} {\bibinfo  {journal} {2D Mater.}\ }\textbf {\bibinfo
  {volume} {3}},\ \bibinfo {pages} {025011} (\bibinfo {year}
  {2016}{\natexlab{b}})}\BibitemShut {NoStop}%
\bibitem [{\citenamefont {Huang}\ \emph {et~al.}(2016)\citenamefont {Huang},
  \citenamefont {Qiao}, \citenamefont {He}, \citenamefont {Bliznakov},
  \citenamefont {Sutter}, \citenamefont {Chen}, \citenamefont {Luo},
  \citenamefont {Meng}, \citenamefont {Su}, \citenamefont {Decker},
  \citenamefont {Ji}, \citenamefont {Ruoff},\ and\ \citenamefont
  {Sutter}}]{huang2016}%
  \BibitemOpen
  \bibfield  {author} {\bibinfo {author} {\bibfnamefont {Y.}~\bibnamefont
  {Huang}}, \bibinfo {author} {\bibfnamefont {J.}~\bibnamefont {Qiao}},
  \bibinfo {author} {\bibfnamefont {K.}~\bibnamefont {He}}, \bibinfo {author}
  {\bibfnamefont {S.}~\bibnamefont {Bliznakov}}, \bibinfo {author}
  {\bibfnamefont {E.}~\bibnamefont {Sutter}}, \bibinfo {author} {\bibfnamefont
  {X.}~\bibnamefont {Chen}}, \bibinfo {author} {\bibfnamefont {D.}~\bibnamefont
  {Luo}}, \bibinfo {author} {\bibfnamefont {F.}~\bibnamefont {Meng}}, \bibinfo
  {author} {\bibfnamefont {D.}~\bibnamefont {Su}}, \bibinfo {author}
  {\bibfnamefont {J.}~\bibnamefont {Decker}}, \bibinfo {author} {\bibfnamefont
  {W.}~\bibnamefont {Ji}}, \bibinfo {author} {\bibfnamefont {R.~S.}\
  \bibnamefont {Ruoff}}, \ and\ \bibinfo {author} {\bibfnamefont
  {P.}~\bibnamefont {Sutter}},\ }\href@noop {} {\bibfield  {journal} {\bibinfo
  {journal} {Chem. Mater.}\ }\textbf {\bibinfo {volume} {28}},\ \bibinfo
  {pages} {8330} (\bibinfo {year} {2016})}\BibitemShut {NoStop}%
\bibitem [{\citenamefont {Dresselhaus}\ and\ \citenamefont
  {Dresselhaus}(1981)}]{Dresselhaus1981}%
  \BibitemOpen
  \bibfield  {author} {\bibinfo {author} {\bibfnamefont {M.~S.}\ \bibnamefont
  {Dresselhaus}}\ and\ \bibinfo {author} {\bibfnamefont {G.}~\bibnamefont
  {Dresselhaus}},\ }\href@noop {} {\bibfield  {journal} {\bibinfo  {journal}
  {Adv. Phys.}\ }\textbf {\bibinfo {volume} {30}},\ \bibinfo {pages} {1}
  (\bibinfo {year} {1981})}\BibitemShut {NoStop}%
\bibitem [{\citenamefont {Keyes}(1953)}]{Keyes1953}%
  \BibitemOpen
  \bibfield  {author} {\bibinfo {author} {\bibfnamefont {R.~W.}\ \bibnamefont
  {Keyes}},\ }\href@noop {} {\bibfield  {journal} {\bibinfo  {journal} {Phys.
  Rev.}\ }\textbf {\bibinfo {volume} {92}},\ \bibinfo {pages} {580} (\bibinfo
  {year} {1953})}\BibitemShut {NoStop}%
\bibitem [{\citenamefont {Warschauer}(1963)}]{Warschauer1963}%
  \BibitemOpen
  \bibfield  {author} {\bibinfo {author} {\bibfnamefont {D.}~\bibnamefont
  {Warschauer}},\ }\href@noop {} {\bibfield  {journal} {\bibinfo  {journal} {J.
  Appl. Phys.}\ }\textbf {\bibinfo {volume} {34}},\ \bibinfo {pages} {1853}
  (\bibinfo {year} {1963})}\BibitemShut {NoStop}%
\bibitem [{\citenamefont {Maruyama}\ \emph {et~al.}(1981)\citenamefont
  {Maruyama}, \citenamefont {Suzuki}, \citenamefont {Kobayashi},\ and\
  \citenamefont {Tanuma}}]{Maruyama1981}%
  \BibitemOpen
  \bibfield  {author} {\bibinfo {author} {\bibfnamefont {Y.}~\bibnamefont
  {Maruyama}}, \bibinfo {author} {\bibfnamefont {S.}~\bibnamefont {Suzuki}},
  \bibinfo {author} {\bibfnamefont {K.}~\bibnamefont {Kobayashi}}, \ and\
  \bibinfo {author} {\bibfnamefont {S.}~\bibnamefont {Tanuma}},\ }\href@noop {}
  {\bibfield  {journal} {\bibinfo  {journal} {Physica B+C}\ }\textbf {\bibinfo
  {volume} {105}},\ \bibinfo {pages} {99} (\bibinfo {year} {1981})}\BibitemShut
  {NoStop}%
\bibitem [{\citenamefont {Harada}\ \emph {et~al.}(1982)\citenamefont {Harada},
  \citenamefont {Murano}, \citenamefont {Shirotani}, \citenamefont
  {Takahashi},\ and\ \citenamefont {Maruyama}}]{Harada1982}%
  \BibitemOpen
  \bibfield  {author} {\bibinfo {author} {\bibfnamefont {Y.}~\bibnamefont
  {Harada}}, \bibinfo {author} {\bibfnamefont {K.}~\bibnamefont {Murano}},
  \bibinfo {author} {\bibfnamefont {I.}~\bibnamefont {Shirotani}}, \bibinfo
  {author} {\bibfnamefont {T.}~\bibnamefont {Takahashi}}, \ and\ \bibinfo
  {author} {\bibfnamefont {Y.}~\bibnamefont {Maruyama}},\ }\href@noop {}
  {\bibfield  {journal} {\bibinfo  {journal} {Solid State Commun.}\ }\textbf
  {\bibinfo {volume} {44}},\ \bibinfo {pages} {877} (\bibinfo {year}
  {1982})}\BibitemShut {NoStop}%
\bibitem [{\citenamefont {Kikegawa}\ and\ \citenamefont
  {Iwasaki}(1983)}]{Akahama1983}%
  \BibitemOpen
  \bibfield  {author} {\bibinfo {author} {\bibfnamefont {T.}~\bibnamefont
  {Kikegawa}}\ and\ \bibinfo {author} {\bibfnamefont {H.}~\bibnamefont
  {Iwasaki}},\ }\href@noop {} {\bibfield  {journal} {\bibinfo  {journal} {J.
  Phys. Soc. Jpn.}\ }\textbf {\bibinfo {volume} {52}},\ \bibinfo {pages} {2148}
  (\bibinfo {year} {1983})}\BibitemShut {NoStop}%
\bibitem [{\citenamefont {Narita}\ \emph {et~al.}(1983)\citenamefont {Narita},
  \citenamefont {Akahama}, \citenamefont {Tsukiyama}, \citenamefont {Muro},
  \citenamefont {Mori}, \citenamefont {Endo}, \citenamefont {Taniguchi},
  \citenamefont {Seki}, \citenamefont {Suga}, \citenamefont {Mikuni},\ and\
  \citenamefont {Kanzaki}}]{Narita1983}%
  \BibitemOpen
  \bibfield  {author} {\bibinfo {author} {\bibfnamefont {S.}~\bibnamefont
  {Narita}}, \bibinfo {author} {\bibfnamefont {Y.}~\bibnamefont {Akahama}},
  \bibinfo {author} {\bibfnamefont {Y.}~\bibnamefont {Tsukiyama}}, \bibinfo
  {author} {\bibfnamefont {K.}~\bibnamefont {Muro}}, \bibinfo {author}
  {\bibfnamefont {S.}~\bibnamefont {Mori}}, \bibinfo {author} {\bibfnamefont
  {S.}~\bibnamefont {Endo}}, \bibinfo {author} {\bibfnamefont {M.}~\bibnamefont
  {Taniguchi}}, \bibinfo {author} {\bibfnamefont {M.}~\bibnamefont {Seki}},
  \bibinfo {author} {\bibfnamefont {S.}~\bibnamefont {Suga}}, \bibinfo {author}
  {\bibfnamefont {A.}~\bibnamefont {Mikuni}}, \ and\ \bibinfo {author}
  {\bibfnamefont {H.}~\bibnamefont {Kanzaki}},\ }\href@noop {} {\bibfield
  {journal} {\bibinfo  {journal} {Physica B+C}\ }\textbf {\bibinfo {volume}
  {117}},\ \bibinfo {pages} {422} (\bibinfo {year} {1983})}\BibitemShut
  {NoStop}%
\bibitem [{\citenamefont {Takahashi}\ \emph {et~al.}(1984)\citenamefont
  {Takahashi}, \citenamefont {Tokailin}, \citenamefont {Suzuki},\ and\
  \citenamefont {Sagawa}}]{Takahashi1984}%
  \BibitemOpen
  \bibfield  {author} {\bibinfo {author} {\bibfnamefont {T.}~\bibnamefont
  {Takahashi}}, \bibinfo {author} {\bibfnamefont {H.}~\bibnamefont {Tokailin}},
  \bibinfo {author} {\bibfnamefont {S.}~\bibnamefont {Suzuki}}, \ and\ \bibinfo
  {author} {\bibfnamefont {T.}~\bibnamefont {Sagawa}},\ }\href@noop {}
  {\bibfield  {journal} {\bibinfo  {journal} {Phys. Rev. B}\ }\textbf {\bibinfo
  {volume} {29}},\ \bibinfo {pages} {1105} (\bibinfo {year}
  {1984})}\BibitemShut {NoStop}%
\bibitem [{\citenamefont {Hayasi}\ \emph {et~al.}(1984)\citenamefont {Hayasi},
  \citenamefont {Takahashi}, \citenamefont {Asahina}, \citenamefont {Sagawa},
  \citenamefont {Morita},\ and\ \citenamefont {Shirotani}}]{Hayashi1984}%
  \BibitemOpen
  \bibfield  {author} {\bibinfo {author} {\bibfnamefont {Y.}~\bibnamefont
  {Hayasi}}, \bibinfo {author} {\bibfnamefont {T.}~\bibnamefont {Takahashi}},
  \bibinfo {author} {\bibfnamefont {H.}~\bibnamefont {Asahina}}, \bibinfo
  {author} {\bibfnamefont {T.}~\bibnamefont {Sagawa}}, \bibinfo {author}
  {\bibfnamefont {A.}~\bibnamefont {Morita}}, \ and\ \bibinfo {author}
  {\bibfnamefont {I.}~\bibnamefont {Shirotani}},\ }\href@noop {} {\bibfield
  {journal} {\bibinfo  {journal} {Phys. Rev. B}\ }\textbf {\bibinfo {volume}
  {30}},\ \bibinfo {pages} {1891} (\bibinfo {year} {1984})}\BibitemShut
  {NoStop}%
\bibitem [{\citenamefont {Baba}\ \emph {et~al.}(1991)\citenamefont {Baba},
  \citenamefont {Izumida}, \citenamefont {Morita}, \citenamefont {Korke},\ and\
  \citenamefont {Fukase}}]{Baba1991_1}%
  \BibitemOpen
  \bibfield  {author} {\bibinfo {author} {\bibfnamefont {M.}~\bibnamefont
  {Baba}}, \bibinfo {author} {\bibfnamefont {F.}~\bibnamefont {Izumida}},
  \bibinfo {author} {\bibfnamefont {A.}~\bibnamefont {Morita}}, \bibinfo
  {author} {\bibfnamefont {Y.}~\bibnamefont {Korke}}, \ and\ \bibinfo {author}
  {\bibfnamefont {T.}~\bibnamefont {Fukase}},\ }\href@noop {} {\bibfield
  {journal} {\bibinfo  {journal} {Jpn. J. Appl. Phys.}\ }\textbf {\bibinfo
  {volume} {30}},\ \bibinfo {pages} {1753} (\bibinfo {year}
  {1991})}\BibitemShut {NoStop}%
\bibitem [{\citenamefont {Han}\ \emph {et~al.}(2014)\citenamefont {Han},
  \citenamefont {Yao}, \citenamefont {Bai}, \citenamefont {Miao}, \citenamefont
  {Zhu}, \citenamefont {Guan}, \citenamefont {Wang}, \citenamefont {Gao},
  \citenamefont {Liu}, \citenamefont {Qian}, \citenamefont {Liu},\ and\
  \citenamefont {Jia}}]{Han2014}%
  \BibitemOpen
  \bibfield  {author} {\bibinfo {author} {\bibfnamefont {C.~Q.}\ \bibnamefont
  {Han}}, \bibinfo {author} {\bibfnamefont {M.~Y.}\ \bibnamefont {Yao}},
  \bibinfo {author} {\bibfnamefont {X.~X.}\ \bibnamefont {Bai}}, \bibinfo
  {author} {\bibfnamefont {L.}~\bibnamefont {Miao}}, \bibinfo {author}
  {\bibfnamefont {F.}~\bibnamefont {Zhu}}, \bibinfo {author} {\bibfnamefont
  {D.~D.}\ \bibnamefont {Guan}}, \bibinfo {author} {\bibfnamefont
  {S.}~\bibnamefont {Wang}}, \bibinfo {author} {\bibfnamefont {C.~L.}\
  \bibnamefont {Gao}}, \bibinfo {author} {\bibfnamefont {C.}~\bibnamefont
  {Liu}}, \bibinfo {author} {\bibfnamefont {D.}~\bibnamefont {Qian}}, \bibinfo
  {author} {\bibfnamefont {Y.}~\bibnamefont {Liu}}, \ and\ \bibinfo {author}
  {\bibfnamefont {J.}~\bibnamefont {Jia}},\ }\href@noop {} {\bibfield
  {journal} {\bibinfo  {journal} {Phys. Rev. B}\ }\textbf {\bibinfo {volume}
  {90}},\ \bibinfo {pages} {085101} (\bibinfo {year} {2014})}\BibitemShut
  {NoStop}%
\bibitem [{\citenamefont {Lange}\ \emph {et~al.}(2007)\citenamefont {Lange},
  \citenamefont {Schmidt},\ and\ \citenamefont {Nilges}}]{Lange2007}%
  \BibitemOpen
  \bibfield  {author} {\bibinfo {author} {\bibfnamefont {S.}~\bibnamefont
  {Lange}}, \bibinfo {author} {\bibfnamefont {P.}~\bibnamefont {Schmidt}}, \
  and\ \bibinfo {author} {\bibfnamefont {T.}~\bibnamefont {Nilges}},\
  }\href@noop {} {\bibfield  {journal} {\bibinfo  {journal} {Inorg. Chem.}\
  }\textbf {\bibinfo {volume} {46}},\ \bibinfo {pages} {4028} (\bibinfo {year}
  {2007})}\BibitemShut {NoStop}%
\bibitem [{\citenamefont {Slichter}(1989)}]{slichter}%
  \BibitemOpen
  \bibfield  {author} {\bibinfo {author} {\bibfnamefont {C.~P.}\ \bibnamefont
  {Slichter}},\ }\href@noop {} {\emph {\bibinfo {title} {{Principles of
  Magnetic Resonance}}}}\ (\bibinfo  {publisher} {Springer},\ \bibinfo
  {address} {New York},\ \bibinfo {year} {1989})\BibitemShut {NoStop}%
\bibitem [{\citenamefont {Fukushima}\ and\ \citenamefont
  {Roeder}(1981)}]{fukushima}%
  \BibitemOpen
  \bibfield  {author} {\bibinfo {author} {\bibfnamefont {E.}~\bibnamefont
  {Fukushima}}\ and\ \bibinfo {author} {\bibfnamefont {S.~B.~W.}\ \bibnamefont
  {Roeder}},\ }\href@noop {} {\emph {\bibinfo {title} {{Experimental Pulse NMR:
  A Nuts and Bolts Approach}}}}\ (\bibinfo  {publisher} {Addison-Wesley
  Publishing Company},\ \bibinfo {address} {Don Mills, Ontario},\ \bibinfo
  {year} {1981})\BibitemShut {NoStop}%
\bibitem [{\citenamefont {Klein}\ \emph {et~al.}(1993)\citenamefont {Klein},
  \citenamefont {Donovan}, \citenamefont {Dressel},\ and\ \citenamefont
  {Gr\"uner}}]{Klein1993}%
  \BibitemOpen
  \bibfield  {author} {\bibinfo {author} {\bibfnamefont {O.}~\bibnamefont
  {Klein}}, \bibinfo {author} {\bibfnamefont {S.}~\bibnamefont {Donovan}},
  \bibinfo {author} {\bibfnamefont {M.}~\bibnamefont {Dressel}}, \ and\
  \bibinfo {author} {\bibfnamefont {G.}~\bibnamefont {Gr\"uner}},\ }\href@noop
  {} {\bibfield  {journal} {\bibinfo  {journal} {Int. J. Infrared Millimeter
  Waves}\ }\textbf {\bibinfo {volume} {14}},\ \bibinfo {pages} {2423} (\bibinfo
  {year} {1993})}\BibitemShut {NoStop}%
\bibitem [{\citenamefont {Donovan}\ \emph {et~al.}(1993)\citenamefont
  {Donovan}, \citenamefont {Klein}, \citenamefont {Dressel}, \citenamefont
  {Holczer},\ and\ \citenamefont {Gr\"uner}}]{Donovan1993}%
  \BibitemOpen
  \bibfield  {author} {\bibinfo {author} {\bibfnamefont {S.}~\bibnamefont
  {Donovan}}, \bibinfo {author} {\bibfnamefont {O.}~\bibnamefont {Klein}},
  \bibinfo {author} {\bibfnamefont {M.}~\bibnamefont {Dressel}}, \bibinfo
  {author} {\bibfnamefont {K.}~\bibnamefont {Holczer}}, \ and\ \bibinfo
  {author} {\bibfnamefont {G.}~\bibnamefont {Gr\"uner}},\ }\href@noop {}
  {\bibfield  {journal} {\bibinfo  {journal} {Int. J. Infrared Millimeter
  Waves}\ }\textbf {\bibinfo {volume} {14}},\ \bibinfo {pages} {2459} (\bibinfo
  {year} {1993})}\BibitemShut {NoStop}%
\bibitem [{\citenamefont {Kitano}\ \emph {et~al.}(2002)\citenamefont {Kitano},
  \citenamefont {Matsuo}, \citenamefont {Miwa}, \citenamefont {Maeda},
  \citenamefont {Takenobu}, \citenamefont {Iwasa},\ and\ \citenamefont
  {Mitani}}]{Kitano2002}%
  \BibitemOpen
  \bibfield  {author} {\bibinfo {author} {\bibfnamefont {H.}~\bibnamefont
  {Kitano}}, \bibinfo {author} {\bibfnamefont {R.}~\bibnamefont {Matsuo}},
  \bibinfo {author} {\bibfnamefont {K.}~\bibnamefont {Miwa}}, \bibinfo {author}
  {\bibfnamefont {A.}~\bibnamefont {Maeda}}, \bibinfo {author} {\bibfnamefont
  {T.}~\bibnamefont {Takenobu}}, \bibinfo {author} {\bibfnamefont
  {Y.}~\bibnamefont {Iwasa}}, \ and\ \bibinfo {author} {\bibfnamefont
  {T.}~\bibnamefont {Mitani}},\ }\href@noop {} {\bibfield  {journal} {\bibinfo
  {journal} {Phys. Rev. Lett.}\ }\textbf {\bibinfo {volume} {88}},\ \bibinfo
  {pages} {096401} (\bibinfo {year} {2002})}\BibitemShut {NoStop}%
\bibitem [{\citenamefont {Sugai}(1985)}]{Sugai1985}%
  \BibitemOpen
  \bibfield  {author} {\bibinfo {author} {\bibfnamefont {S.}~\bibnamefont
  {Sugai}},\ }\href@noop {} {\bibfield  {journal} {\bibinfo  {journal} {Solid
  State Commun.}\ }\textbf {\bibinfo {volume} {53}},\ \bibinfo {pages} {753}
  (\bibinfo {year} {1985})}\BibitemShut {NoStop}%
\bibitem [{\citenamefont {Vanderborgh}\ and\ \citenamefont
  {Schifer}(1989)}]{Vanderborgh1989}%
  \BibitemOpen
  \bibfield  {author} {\bibinfo {author} {\bibfnamefont {C.~A.}\ \bibnamefont
  {Vanderborgh}}\ and\ \bibinfo {author} {\bibfnamefont {D.}~\bibnamefont
  {Schifer}},\ }\href@noop {} {\bibfield  {journal} {\bibinfo  {journal} {Phys.
  Rev. B}\ }\textbf {\bibinfo {volume} {40}},\ \bibinfo {pages} {9595}
  (\bibinfo {year} {1989})}\BibitemShut {NoStop}%
\bibitem [{\citenamefont {Akahama}\ \emph {et~al.}(1997)\citenamefont
  {Akahama}, \citenamefont {Kobayashi},\ and\ \citenamefont
  {Kawamura}}]{Akahama1997}%
  \BibitemOpen
  \bibfield  {author} {\bibinfo {author} {\bibfnamefont {Y.}~\bibnamefont
  {Akahama}}, \bibinfo {author} {\bibfnamefont {M.}~\bibnamefont {Kobayashi}},
  \ and\ \bibinfo {author} {\bibfnamefont {H.}~\bibnamefont {Kawamura}},\
  }\href@noop {} {\bibfield  {journal} {\bibinfo  {journal} {Solid State
  Commun.}\ }\textbf {\bibinfo {volume} {104}},\ \bibinfo {pages} {311}
  (\bibinfo {year} {1997})}\BibitemShut {NoStop}%
\bibitem [{\citenamefont {Late}(2015)}]{Late2015}%
  \BibitemOpen
  \bibfield  {author} {\bibinfo {author} {\bibfnamefont {D.~J.}\ \bibnamefont
  {Late}},\ }\href@noop {} {\bibfield  {journal} {\bibinfo  {journal} {ACS
  Appl. Mater. Interfaces}\ }\textbf {\bibinfo {volume} {7}},\ \bibinfo {pages}
  {5857} (\bibinfo {year} {2015})}\BibitemShut {NoStop}%
\bibitem [{\citenamefont {Ribeiro}\ \emph {et~al.}(2015)\citenamefont
  {Ribeiro}, \citenamefont {Pimenta}, \citenamefont {{{de~Matos}}},
  \citenamefont {Moreira}, \citenamefont {Rodin}, \citenamefont {Zapata},
  \citenamefont {{{de~Souza}}},\ and\ \citenamefont
  {{{Castro~Neto}}}}]{Riberio2015}%
  \BibitemOpen
  \bibfield  {author} {\bibinfo {author} {\bibfnamefont {H.~B.}\ \bibnamefont
  {Ribeiro}}, \bibinfo {author} {\bibfnamefont {M.~A.}\ \bibnamefont
  {Pimenta}}, \bibinfo {author} {\bibfnamefont {C.~J.~S.}\ \bibnamefont
  {{{de~Matos}}}}, \bibinfo {author} {\bibfnamefont {R.~L.}\ \bibnamefont
  {Moreira}}, \bibinfo {author} {\bibfnamefont {A.~S.}\ \bibnamefont {Rodin}},
  \bibinfo {author} {\bibfnamefont {J.~D.}\ \bibnamefont {Zapata}}, \bibinfo
  {author} {\bibfnamefont {E.~A.~T.}\ \bibnamefont {{{de~Souza}}}}, \ and\
  \bibinfo {author} {\bibfnamefont {A.~H.}\ \bibnamefont {{{Castro~Neto}}}},\
  }\href@noop {} {\bibfield  {journal} {\bibinfo  {journal} {ACS Nano}\
  }\textbf {\bibinfo {volume} {9}},\ \bibinfo {pages} {4270} (\bibinfo {year}
  {2015})}\BibitemShut {NoStop}%
\bibitem [{\citenamefont {Ling}\ \emph {et~al.}(2016)\citenamefont {Ling},
  \citenamefont {Huang}, \citenamefont {Hasdeo}, \citenamefont {Liang},
  \citenamefont {Parkin}, \citenamefont {Tatsumi}, \citenamefont {Nugraha},
  \citenamefont {Puretzky}, \citenamefont {Das}, \citenamefont {Sumpter},
  \citenamefont {Geohegan}, \citenamefont {Kong}, \citenamefont {Saito},
  \citenamefont {Drndic}, \citenamefont {Meunier},\ and\ \citenamefont
  {Dresselhaus}}]{Ling2016}%
  \BibitemOpen
  \bibfield  {author} {\bibinfo {author} {\bibfnamefont {X.}~\bibnamefont
  {Ling}}, \bibinfo {author} {\bibfnamefont {S.}~\bibnamefont {Huang}},
  \bibinfo {author} {\bibfnamefont {E.~H.}\ \bibnamefont {Hasdeo}}, \bibinfo
  {author} {\bibfnamefont {L.}~\bibnamefont {Liang}}, \bibinfo {author}
  {\bibfnamefont {W.~M.}\ \bibnamefont {Parkin}}, \bibinfo {author}
  {\bibfnamefont {Y.}~\bibnamefont {Tatsumi}}, \bibinfo {author} {\bibfnamefont
  {A.~R.~T.}\ \bibnamefont {Nugraha}}, \bibinfo {author} {\bibfnamefont
  {A.~A.}\ \bibnamefont {Puretzky}}, \bibinfo {author} {\bibfnamefont {P.~M.}\
  \bibnamefont {Das}}, \bibinfo {author} {\bibfnamefont {B.~G.}\ \bibnamefont
  {Sumpter}}, \bibinfo {author} {\bibfnamefont {D.~B.}\ \bibnamefont
  {Geohegan}}, \bibinfo {author} {\bibfnamefont {J.}~\bibnamefont {Kong}},
  \bibinfo {author} {\bibfnamefont {R.}~\bibnamefont {Saito}}, \bibinfo
  {author} {\bibfnamefont {M.}~\bibnamefont {Drndic}}, \bibinfo {author}
  {\bibfnamefont {V.}~\bibnamefont {Meunier}}, \ and\ \bibinfo {author}
  {\bibfnamefont {M.~S.}\ \bibnamefont {Dresselhaus}},\ }\href@noop {}
  {\bibfield  {journal} {\bibinfo  {journal} {Nano Lett.}\ }\textbf {\bibinfo
  {volume} {16}},\ \bibinfo {pages} {2260} (\bibinfo {year}
  {2016})}\BibitemShut {NoStop}%
\bibitem [{\citenamefont {Greenwood}\ and\ \citenamefont
  {Earnshaw}(1998)}]{greenwood1998}%
  \BibitemOpen
  \bibfield  {author} {\bibinfo {author} {\bibfnamefont {N.~N.}\ \bibnamefont
  {Greenwood}}\ and\ \bibinfo {author} {\bibfnamefont {A.}~\bibnamefont
  {Earnshaw}},\ }\href@noop {} {\emph {\bibinfo {title} {{Chemistry of the
  Elements}}}}\ (\bibinfo  {publisher} {Butterworth},\ \bibinfo {address}
  {London, United Kingdom},\ \bibinfo {year} {1998})\BibitemShut {NoStop}%
\bibitem [{\citenamefont {Andrew}\ and\ \citenamefont
  {Wynn}(1966)}]{andrew1966}%
  \BibitemOpen
  \bibfield  {author} {\bibinfo {author} {\bibfnamefont {E.~R.}\ \bibnamefont
  {Andrew}}\ and\ \bibinfo {author} {\bibfnamefont {V.~T.}\ \bibnamefont
  {Wynn}},\ }\href@noop {} {\bibfield  {journal} {\bibinfo  {journal}
  {Proceedings of the Royal Society of London A: Mathematical, Physical and
  Engineering Sciences}\ }\textbf {\bibinfo {volume} {291}},\ \bibinfo {pages}
  {257} (\bibinfo {year} {1966})}\BibitemShut {NoStop}%
\bibitem [{\citenamefont {Morgan}\ and\ \citenamefont
  {Wazer}(1975)}]{morgan1975}%
  \BibitemOpen
  \bibfield  {author} {\bibinfo {author} {\bibfnamefont {W.~E.}\ \bibnamefont
  {Morgan}}\ and\ \bibinfo {author} {\bibfnamefont {J.~R.~V.}\ \bibnamefont
  {Wazer}},\ }\href@noop {} {\bibfield  {journal} {\bibinfo  {journal} {‎J.
  Am. Chem. Soc.}\ }\textbf {\bibinfo {volume} {97}},\ \bibinfo {pages} {6347}
  (\bibinfo {year} {1975})}\BibitemShut {NoStop}%
\bibitem [{\citenamefont {Simon}\ \emph {et~al.}(2006)\citenamefont {Simon},
  \citenamefont {Kuzmany}, \citenamefont {N\'{a}fr\'{a}di}, \citenamefont
  {Feh\'{e}r}, \citenamefont {Forr\'{o}}, \citenamefont {F\"{u}l\"{o}p},
  \citenamefont {J\'{a}nossy}, \citenamefont {Korecz}, \citenamefont
  {Rockenbauer}, \citenamefont {Hauke},\ and\ \citenamefont
  {Hirsch}}]{simon2006}%
  \BibitemOpen
  \bibfield  {author} {\bibinfo {author} {\bibfnamefont {F.}~\bibnamefont
  {Simon}}, \bibinfo {author} {\bibfnamefont {H.}~\bibnamefont {Kuzmany}},
  \bibinfo {author} {\bibfnamefont {B.}~\bibnamefont {N\'{a}fr\'{a}di}},
  \bibinfo {author} {\bibfnamefont {T.}~\bibnamefont {Feh\'{e}r}}, \bibinfo
  {author} {\bibfnamefont {L.}~\bibnamefont {Forr\'{o}}}, \bibinfo {author}
  {\bibfnamefont {F.}~\bibnamefont {F\"{u}l\"{o}p}}, \bibinfo {author}
  {\bibfnamefont {A.}~\bibnamefont {J\'{a}nossy}}, \bibinfo {author}
  {\bibfnamefont {L.}~\bibnamefont {Korecz}}, \bibinfo {author} {\bibfnamefont
  {A.}~\bibnamefont {Rockenbauer}}, \bibinfo {author} {\bibfnamefont
  {F.}~\bibnamefont {Hauke}}, \ and\ \bibinfo {author} {\bibfnamefont
  {A.}~\bibnamefont {Hirsch}},\ }\href@noop {} {\bibfield  {journal} {\bibinfo
  {journal} {Phys. Rev. Lett.}\ }\textbf {\bibinfo {volume} {97}},\ \bibinfo
  {pages} {136801} (\bibinfo {year} {2006})}\BibitemShut {NoStop}%
\bibitem [{\citenamefont {Simon}\ \emph {et~al.}(2007)\citenamefont {Simon},
  \citenamefont {Quintavalle}, \citenamefont {J\'{a}nossy}, \citenamefont
  {N\'{a}fr\'{a}di}, \citenamefont {Forr\'{o}}, \citenamefont {Kuzmany},
  \citenamefont {Hauke}, \citenamefont {Hirsch}, \citenamefont {Mende},\ and\
  \citenamefont {Mehring}}]{simon2007}%
  \BibitemOpen
  \bibfield  {author} {\bibinfo {author} {\bibfnamefont {F.}~\bibnamefont
  {Simon}}, \bibinfo {author} {\bibfnamefont {D.}~\bibnamefont {Quintavalle}},
  \bibinfo {author} {\bibfnamefont {A.}~\bibnamefont {J\'{a}nossy}}, \bibinfo
  {author} {\bibfnamefont {B.}~\bibnamefont {N\'{a}fr\'{a}di}}, \bibinfo
  {author} {\bibfnamefont {L.}~\bibnamefont {Forr\'{o}}}, \bibinfo {author}
  {\bibfnamefont {H.}~\bibnamefont {Kuzmany}}, \bibinfo {author} {\bibfnamefont
  {F.}~\bibnamefont {Hauke}}, \bibinfo {author} {\bibfnamefont
  {A.}~\bibnamefont {Hirsch}}, \bibinfo {author} {\bibfnamefont
  {J.}~\bibnamefont {Mende}}, \ and\ \bibinfo {author} {\bibfnamefont
  {M.}~\bibnamefont {Mehring}},\ }\href@noop {} {\bibfield  {journal} {\bibinfo
   {journal} {phys. stat. sol. (b)}\ }\textbf {\bibinfo {volume} {244}},\
  \bibinfo {pages} {3885} (\bibinfo {year} {2007})}\BibitemShut {NoStop}%
\bibitem [{\citenamefont {Quintavalle}\ \emph {et~al.}(2009)\citenamefont
  {Quintavalle}, \citenamefont {Simon}, \citenamefont {Klupp}, \citenamefont
  {Kiss}, \citenamefont {Bortel}, \citenamefont {Pekker},\ and\ \citenamefont
  {J\'{a}nossy}}]{quintavalle2009}%
  \BibitemOpen
  \bibfield  {author} {\bibinfo {author} {\bibfnamefont {D.}~\bibnamefont
  {Quintavalle}}, \bibinfo {author} {\bibfnamefont {F.}~\bibnamefont {Simon}},
  \bibinfo {author} {\bibfnamefont {G.}~\bibnamefont {Klupp}}, \bibinfo
  {author} {\bibfnamefont {L.~F.}\ \bibnamefont {Kiss}}, \bibinfo {author}
  {\bibfnamefont {G.}~\bibnamefont {Bortel}}, \bibinfo {author} {\bibfnamefont
  {S.}~\bibnamefont {Pekker}}, \ and\ \bibinfo {author} {\bibfnamefont
  {A.}~\bibnamefont {J\'{a}nossy}},\ }\href@noop {} {\bibfield  {journal}
  {\bibinfo  {journal} {Phys. Rev. B}\ }\textbf {\bibinfo {volume} {80}},\
  \bibinfo {pages} {033403} (\bibinfo {year} {2009})}\BibitemShut {NoStop}%
\bibitem [{\citenamefont {Fabi\'{a}n}\ \emph {et~al.}(2012)\citenamefont
  {Fabi\'{a}n}, \citenamefont {D\'{o}ra}, \citenamefont {Antal}, \citenamefont
  {Szolnoki}, \citenamefont {Korecz}, \citenamefont {Rockenbauer},
  \citenamefont {Nemes}, \citenamefont {Forr\'{o}},\ and\ \citenamefont
  {Simon}}]{fabian2012}%
  \BibitemOpen
  \bibfield  {author} {\bibinfo {author} {\bibfnamefont {G.}~\bibnamefont
  {Fabi\'{a}n}}, \bibinfo {author} {\bibfnamefont {B.}~\bibnamefont
  {D\'{o}ra}}, \bibinfo {author} {\bibfnamefont {{\'{A}}.}~\bibnamefont
  {Antal}}, \bibinfo {author} {\bibfnamefont {L.}~\bibnamefont {Szolnoki}},
  \bibinfo {author} {\bibfnamefont {L.}~\bibnamefont {Korecz}}, \bibinfo
  {author} {\bibfnamefont {A.}~\bibnamefont {Rockenbauer}}, \bibinfo {author}
  {\bibfnamefont {N.~M.}\ \bibnamefont {Nemes}}, \bibinfo {author}
  {\bibfnamefont {L.}~\bibnamefont {Forr\'{o}}}, \ and\ \bibinfo {author}
  {\bibfnamefont {F.}~\bibnamefont {Simon}},\ }\href@noop {} {\bibfield
  {journal} {\bibinfo  {journal} {Phys. Rev. B}\ }\textbf {\bibinfo {volume}
  {85}},\ \bibinfo {pages} {235405} (\bibinfo {year} {2012})}\BibitemShut
  {NoStop}%
\bibitem [{\citenamefont {M\'{a}rkus}\ \emph {et~al.}(2015)\citenamefont
  {M\'{a}rkus}, \citenamefont {Simon}, \citenamefont {Chac\'{o}n-Torres},
  \citenamefont {Reich}, \citenamefont {Szirmai}, \citenamefont
  {N\'{a}fr\'{a}di}, \citenamefont {Forr\'{o}}, \citenamefont {Pichler},
  \citenamefont {Vecera}, \citenamefont {Hauke},\ and\ \citenamefont
  {Hirsch}}]{markus2015}%
  \BibitemOpen
  \bibfield  {author} {\bibinfo {author} {\bibfnamefont {B.~G.}\ \bibnamefont
  {M\'{a}rkus}}, \bibinfo {author} {\bibfnamefont {F.}~\bibnamefont {Simon}},
  \bibinfo {author} {\bibfnamefont {J.~C.}\ \bibnamefont {Chac\'{o}n-Torres}},
  \bibinfo {author} {\bibfnamefont {S.}~\bibnamefont {Reich}}, \bibinfo
  {author} {\bibfnamefont {P.}~\bibnamefont {Szirmai}}, \bibinfo {author}
  {\bibfnamefont {B.}~\bibnamefont {N\'{a}fr\'{a}di}}, \bibinfo {author}
  {\bibfnamefont {L.}~\bibnamefont {Forr\'{o}}}, \bibinfo {author}
  {\bibfnamefont {T.}~\bibnamefont {Pichler}}, \bibinfo {author} {\bibfnamefont
  {P.}~\bibnamefont {Vecera}}, \bibinfo {author} {\bibfnamefont
  {F.}~\bibnamefont {Hauke}}, \ and\ \bibinfo {author} {\bibfnamefont
  {A.}~\bibnamefont {Hirsch}},\ }\href@noop {} {\bibfield  {journal} {\bibinfo
  {journal} {phys. stat. sol. (b)}\ }\textbf {\bibinfo {volume} {252}},\
  \bibinfo {pages} {2438} (\bibinfo {year} {2015})}\BibitemShut {NoStop}%
\bibitem [{\citenamefont {Maruyama}\ \emph {et~al.}(1986)\citenamefont
  {Maruyama}, \citenamefont {Suzuki}, \citenamefont {Osaki}, \citenamefont
  {Yamaguchi}, \citenamefont {Sakai}, \citenamefont {Nagasato},\ and\
  \citenamefont {Shirotani}}]{Maruyama1986}%
  \BibitemOpen
  \bibfield  {author} {\bibinfo {author} {\bibfnamefont {Y.}~\bibnamefont
  {Maruyama}}, \bibinfo {author} {\bibfnamefont {S.}~\bibnamefont {Suzuki}},
  \bibinfo {author} {\bibfnamefont {T.}~\bibnamefont {Osaki}}, \bibinfo
  {author} {\bibfnamefont {H.}~\bibnamefont {Yamaguchi}}, \bibinfo {author}
  {\bibfnamefont {S.}~\bibnamefont {Sakai}}, \bibinfo {author} {\bibfnamefont
  {K.}~\bibnamefont {Nagasato}}, \ and\ \bibinfo {author} {\bibfnamefont
  {I.}~\bibnamefont {Shirotani}},\ }\href@noop {} {\bibfield  {journal}
  {\bibinfo  {journal} {Bull. Chem. Soc. Jpn.}\ }\textbf {\bibinfo {volume}
  {59}},\ \bibinfo {pages} {1067} (\bibinfo {year} {1986})}\BibitemShut
  {NoStop}%
\bibitem [{\citenamefont {Baba}\ \emph {et~al.}(1989)\citenamefont {Baba},
  \citenamefont {Izumida}, \citenamefont {Takeda},\ and\ \citenamefont
  {Morita}}]{Baba1989_1}%
  \BibitemOpen
  \bibfield  {author} {\bibinfo {author} {\bibfnamefont {M.}~\bibnamefont
  {Baba}}, \bibinfo {author} {\bibfnamefont {F.}~\bibnamefont {Izumida}},
  \bibinfo {author} {\bibfnamefont {Y.}~\bibnamefont {Takeda}}, \ and\ \bibinfo
  {author} {\bibfnamefont {A.}~\bibnamefont {Morita}},\ }\href@noop {}
  {\bibfield  {journal} {\bibinfo  {journal} {Jpn. J. Appl. Phys.}\ }\textbf
  {\bibinfo {volume} {28}},\ \bibinfo {pages} {1019} (\bibinfo {year}
  {1989})}\BibitemShut {NoStop}%
\end{thebibliography}%

\end{document}